\documentclass[fleqn,usenatbib]{mnras}

\usepackage{cancel}
\usepackage[dvipsnames]{xcolor}
\usepackage{ae,aecompl}
\usepackage{natbib}
\usepackage{graphicx}	
\usepackage{amsmath}	
\usepackage{amssymb}	

\usepackage{newtxtext,newtxmath}
\usepackage[T1]{fontenc}
\DeclareRobustCommand{\VAN}[3]{#2}
\let\VANthebibliography\thebibliography
\def\thebibliography{\DeclareRobustCommand{\VAN}[3]{##3}\VANthebibliography}

\title{Substructure Detection Reanalyzed: \\Dark Perturber shown to be a Line-of-Sight Halo}

\author[A. Ç. Şengül et al.]{
Atınç Çağan Şengül$^{1}$,\thanks{sengul@g.harvard.edu}
Cora Dvorkin$^{1},$\thanks{cdvorkin@g.harvard.edu}
Bryan Ostdiek$^{1}$,
and Arthur Tsang$^{1}$\\
$^{1}$Harvard University, Department of Physics, Cambridge, Massachusetts, 02138, U.S.A.
}

\pubyear{2022}
\hypersetup{final}
\begin{document}
\label{firstpage}
\pagerange{\pageref{firstpage}--\pageref{lastpage}}
\maketitle

\begin{abstract}
    Observations of structure at sub-galactic scales are crucial for probing the properties of dark matter, which is the dominant source of gravity in the universe. It will become increasingly important for future surveys to distinguish between line-of-sight halos and subhalos to avoid wrong inferences on the nature of dark matter. We reanalyze a sub-galactic structure (in lens JVAS B1938+666) that has been previously found using the gravitational imaging technique in galaxy-galaxy lensing systems. This structure has been assumed to be a satellite in the halo of the main lens galaxy. We fit the redshift of the perturber of the system as a free parameter, using the multi-plane thin-lens approximation, and find that the redshift of the perturber is $z_\mathrm{int} = 1.42\substack{+0.10 \\ -0.15}$ (with a main lens redshift of $z=0.881$). Our analysis indicates that this structure is more massive than the previous result by an order of magnitude. This constitutes the first dark perturber shown to be a line-of-sight halo with a gravitational lensing method. 
\end{abstract}


\section{Introduction}

We have long known that the matter content of the universe is dominated by dark matter (DM),  whose fundamental nature strongly affects structure formation. Therefore, measuring the matter distribution in the universe can be used to shed light on the properties of DM. 
Measurements of the distribution of galaxies \citep{sdss2019}, weak lensing \citep{DES2021}, and cosmic microwave background fluctuations \citep{planck2018} have shown that the matter distribution at galactic and super-galactic mass scales is consistent with a cold dark matter (CDM) model.
Therefore, in general, the DM models that remain untested are the ones that have predictions that differ from CDM at sub-galactic mass scales  \citep{hm_fit,dm_mw_sim}. Strong gravitational lensing is one powerful probe of such lower-mass scales to test the nature of DM.

Strong gravitational lensing involves a main deflector, which we will refer to as main lens, that bends the light that a background galaxy emits. This results in multiple highly distorted images of the background source galaxy. Small perturbations in the mass distribution of the main deflector cause changes in the pixel brightnesses of the lensed images, which are used to detect such structures. So far, two such perturbers have been detected with this method on \textit{Hubble Space Telescope} (HST) data with claimed masses of $3.5 \times 10^{9} \mathrm{M}_\odot$ and $1.9 \times 10^{8} \mathrm{M}_\odot$ \citep{2009detection,2012Natur}. The second system was also observed with an adaptive optics system mounted on the Keck telescope. Another perturber with mass $1.0 \times 10^{9} \mathrm{M}_\odot$ was detected using interferometry data from Atacama Large Millimeter/submillimeter Array (ALMA) \citep{hezaveh_sub}. These masses were inferred by assuming that the detected structures are subhalos of the main lenses of these lensing systems and, therefore, are tidally stripped by the gravitational pull of the host halo. This tidal stripping strongly depends on how close the subhalo is to the center of the host halo, which is a large source of uncertainty in determining the masses of these substructures. 

Furthermore, it is possible that these perturbers are not associated with the main lens and are halos that lie along the line of sight; we will call this type of perturber ``interlopers''. 
We have previously calculated the expected 2D number densities of subhalos and interlopers for typical lensing geometries and found that most perturbers should be interlopers \citep{int_pow_spec} (in agreement with previous works \citep{Li:2016afu,despali2017}). The effect of interlopers is not identical to
that of the subhalos in terms of perturbations on strong lensing arcs. 
Furthermore, the mass function, which is the number density of halos or subhalos per comoving volume per mass, differs significantly between subhalos and interlopers since they are subject to different physical environments \citep{halo_subhalo_mass_func}. As current and future surveys are expected to grow the number of known lensing systems by several orders of magnitude, we can expect the number of perturbers that are detected gravitationally to grow as well \citep{DES_lensing_cnn1,DESI_lensing_cat}. 
Therefore, we will soon be able to use perturbers to put tight constraints on the halo and subhalo mass functions and, ultimately, to obtain an unbiased inference on the nature of dark matter. This analysis, however, will rely on our ability to distinguish the perturbers as interlopers or subhalos.

In this work, we reanalyzed the system JVAS B1938+666. The data are publicly available on \url{https://mast.stsci.edu/portal/Mashup/Clients/Mast/Portal.html}. This system is one of the two examples, so far, of substructure detection using the gravitational imaging technique \citep{grav_imaging}. We give details about the system below.

\begin{figure}
    \centering
    \includegraphics[width=\linewidth]{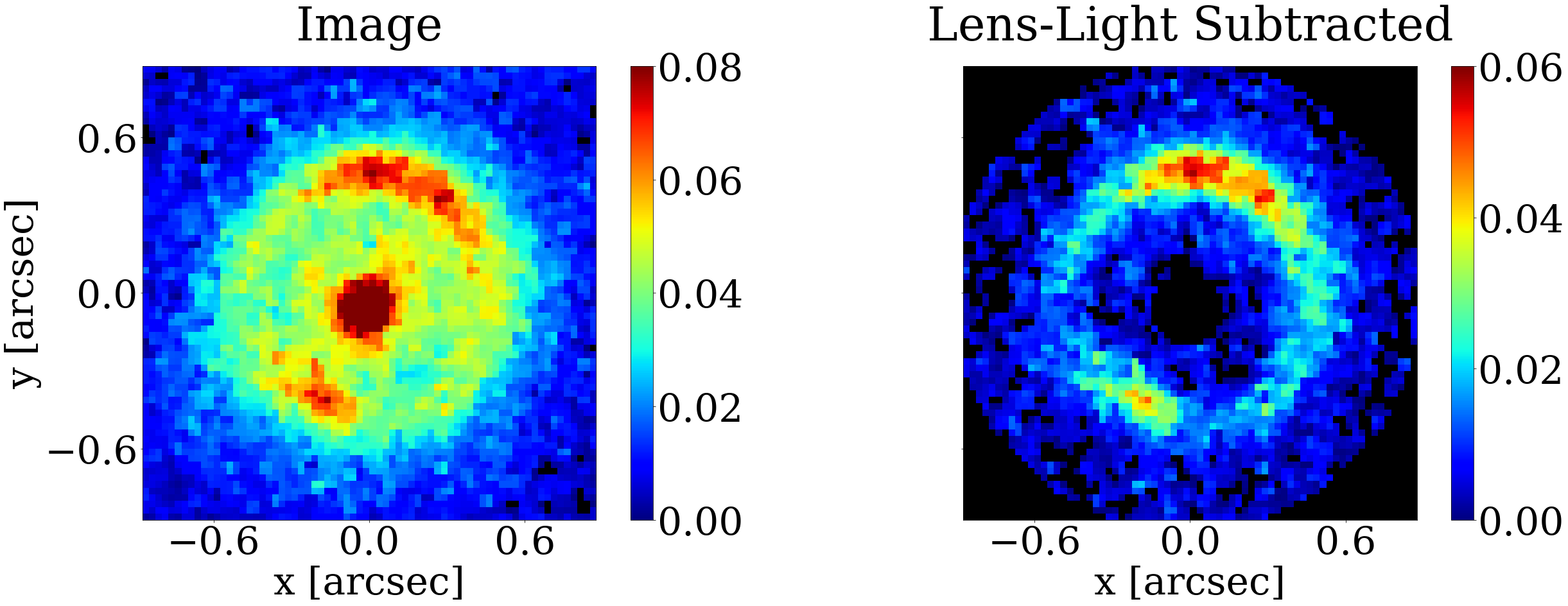}
    \caption{HST image of JVAS B1938+666 used in this study. \textit{Left: }70x70 pixels, 1.6 $\mu$m image. The Einstein ring at  $\theta_E \approx 0.5''$ is visible with its two components. \textit{Right: }The image after lens-light subtraction and masking.} 
    \label{fig:system_raw}
\end{figure}

\section{Methods}
\subsection{JVAS B1938+666}

The system has a main lens at redshift $z_\mathrm{lens} = 0.881$ \citep{JVAS_lens_redshift} and a source galaxy at $z_s = 2.059$ \citep{JVAS_source_redshift}, with an Einstein radius $\theta_E \approx 0.45''$. The image was taken with the Near Infrared Camera and Multi-Object
Spectrograph (NICMOS) on the HST with a 1.6 $\mu$m filter with 6976 seconds of exposure. The pixel size of the {\it drizzled} image is $0.025''$. A dark substructure with mass $M_\mathrm{sub} = (1.9 \pm 0.2) \times 10^{8} \mathrm{M}_\odot$ was detected in previous work at a $12\sigma$ significance using the gravitational imaging method \citep{2012Natur}. For our study, we select a 70$\times$70 pixel region ($ 1.75''\times 1.75''$) centered roughly on the lens light. After lens-light subtraction (see Section~\ref{sec:lenslight} for details), we mask the pixels with radius $r<0.137''$ and $r>0.901''$ from the center of the image since these pixels contain almost no source light, as shown in Fig. \ref{fig:system_raw}.

\subsection{Main Lens and Perturber}
We analyze the image in this work using \texttt{lenstronomy} \citep{lenstronomy1,lenstronomy2}, a publicly available \texttt{Python} package for gravitational lensing. The code used for our analysis is available at \url{https://github.com/acagansengul/interlopers_with_lenstronomy}. The lens model consists of a main lens with a power-law elliptical mass distribution (PEMD) profile \citep{barkana_spemd}, a perturber with a Singular Isothermal Sphere (SIS), Navarro-Frenk-White (NFW) \citep{NFW}, or a pseudo-Jaffe profile , and an external shear. The $x$ and $y$ components of the external shear are denoted $\Gamma_1$ and $\Gamma_2$. The PEMD, SIS, pseudo-Jaffe, and NFW convergences are given by
\begin{align}
    \kappa_\mathrm{PEMD}(\mathbf x) &= \frac{A}{\left[x_1^2 + x_2^2/(1-e^2)\right]^{(\gamma-1)/2}}, \\ \kappa_\mathrm{SIS}(\mathbf x) &= \frac{b_\mathrm{int}}{2|\mathbf x|},\label{eq:profiles}\\ \kappa_\mathrm{pJaffe}(\mathbf{x}) &= \frac{b_\mathrm{int}}{2}\left[\frac{1}{|\mathbf {x}|}- \frac{1}{(|\mathbf x|^2 + r_t^2)^{1/2}}\right], \\ \kappa_\mathrm{NFW}(\mathbf x) &= \frac{2\rho_s r_s}{\Sigma_\mathrm{c}} \frac{1-f\left[(D_l/r_s)\mathbf x\right]}{\left[(D_l/r_s)\mathbf{x}\right]^2 - 1},\label{eq:profiles2}
\end{align}
where the ellipticity orientation for the PEMD is chosen to be along the $x_2$ axis for simplicity, $e$ is the eccentricity, $\gamma$ is the logarithmic slope of the 3D matter distribution, $A = [(3-\gamma)/2][\theta_E^2/\{(1-e)/(1+e)\})]^{(\gamma-1)/2}$ is the amplitude, which reduces to an SIS with $\theta_E = b_\mathrm{int}$ when $e=0$ and $\gamma=2$, $b_\mathrm{int}$ is the lensing strength of the SIS or pseudo-Jaffe, $r_t$ is the pseudo-Jaffe truncation radius, $\rho_s$ is the density of NFW at scale radius $r_s$, $\Sigma_\mathrm{c}$ is the critical surface density, $D_l$ is the angular diameter distance to the perturber, and the function $f(y)$ is given by
\begin{equation}
    f(y) \equiv \left\{
     \begin{array}{@{}l@{\thinspace}l}
       \frac{1}{\sqrt{y^2-1}}\tan^{-1}\sqrt{y^2-1} \quad &: y > 1\\
       \frac{1}{\sqrt{1-y^2}}\tanh^{-1}\sqrt{1-y^2} \quad &: y < 1 \\
       1 \quad &: y = 1\\
     \end{array}
   \right.
   .
\end{equation}
The position parameters $x_\mathrm{int}$ and $y_\mathrm{int}$ are {\it apparent} positions for background interlopers, described in Eq. \eqref{eq:lenseq2} in Appendix \ref{sec:line-of-sight}, meaning that they are the positions at which the interloper would appear due to the lensing of the main lens, if it were luminous. Using true angular positions leads to a strong degeneracy with the redshift of the interloper \citep{DES:2019fny}. This can be intuitively understood by noticing that a background interloper that perturbs the images formed near the Einstein radius needs to be located closer to the central axis connecting the observer to the source as $z_\mathrm{int}$ increases. 

This degeneracy is not perfect,  however. If it were so, measuring the redshift of a perturber would not be possible. The deflection angles caused by the interloper form a vector field with a nonvanishing curl which cannot be recreated by a single thin lensing plane. Moreover, the combined convergence of an interloper and a main lens differs from a naive sum of the two (see Appendix \ref{sec:line-of-sight} for a more detailed discussion). These differences cause slight changes in the pixel brightnesses near the Einstein radius, which is what it is used to constrain the redshift of the perturber.

The NFW profile can be expressed in terms of $M_{200}$ and $c_{200}$ where $c_{200} \equiv r_{200}/r_s$. $c_{200}$ is called \textit{concentration}, and $r_{200}$ is the radius within which the average density is 200 times the critical density of the universe. We will assume the concentration to be a free parameter in our analysis.

\subsection{Lens-Light Subtraction}\label{sec:lenslight}
We use two Sersic profiles for the main lens light and, for the purpose of lens-light subtraction, for the source light as well. The Sersic profiles are given by
\begin{equation}\label{eq:sersic}
    I(\mathbf x) = I_0 \exp\left[-k\left\{\left(\frac{\sqrt{x_1^2 + x_2^2/q^2}}{R_\mathrm{s}}\right)^{1/n_\mathrm{s}} - 1\right\}\right],
\end{equation}
where $R_\mathrm{s}$ is the Sersic radius, $I_0$ is the surface brightness at $R_\mathrm{s}$, $q$ is the ellipticity axis ratio, $n_\mathrm{s}$ is the Sersic index, and $k$ is a normalizing constant that makes $R_\mathrm{s}$ the half-light radius. The centroids of the two lens-light Sersic profiles are fixed to the centroid of the main lens potential. We obtain a best fit for a model that includes a main lens, external shear, two Sersic lens-light components and two Sersic source-light components. The lens-light contribution in this best fit is then subtracted from the image. The Poisson noise is calculated using the pixel values before the lens-light subtraction to capture the noise of the lens-light that remains in the data after lens-light subtraction. The resulting lens-light subtracted image, shown in Fig. \ref{fig:system_raw}, is then masked (as explained in the Results section) and modeled with no lens-light component. The source is modeled with shapelets (as explained below). 

\subsection{Source-Light Regularization and Masking}

\begin{figure}
    \centering
    \includegraphics[width=\linewidth]{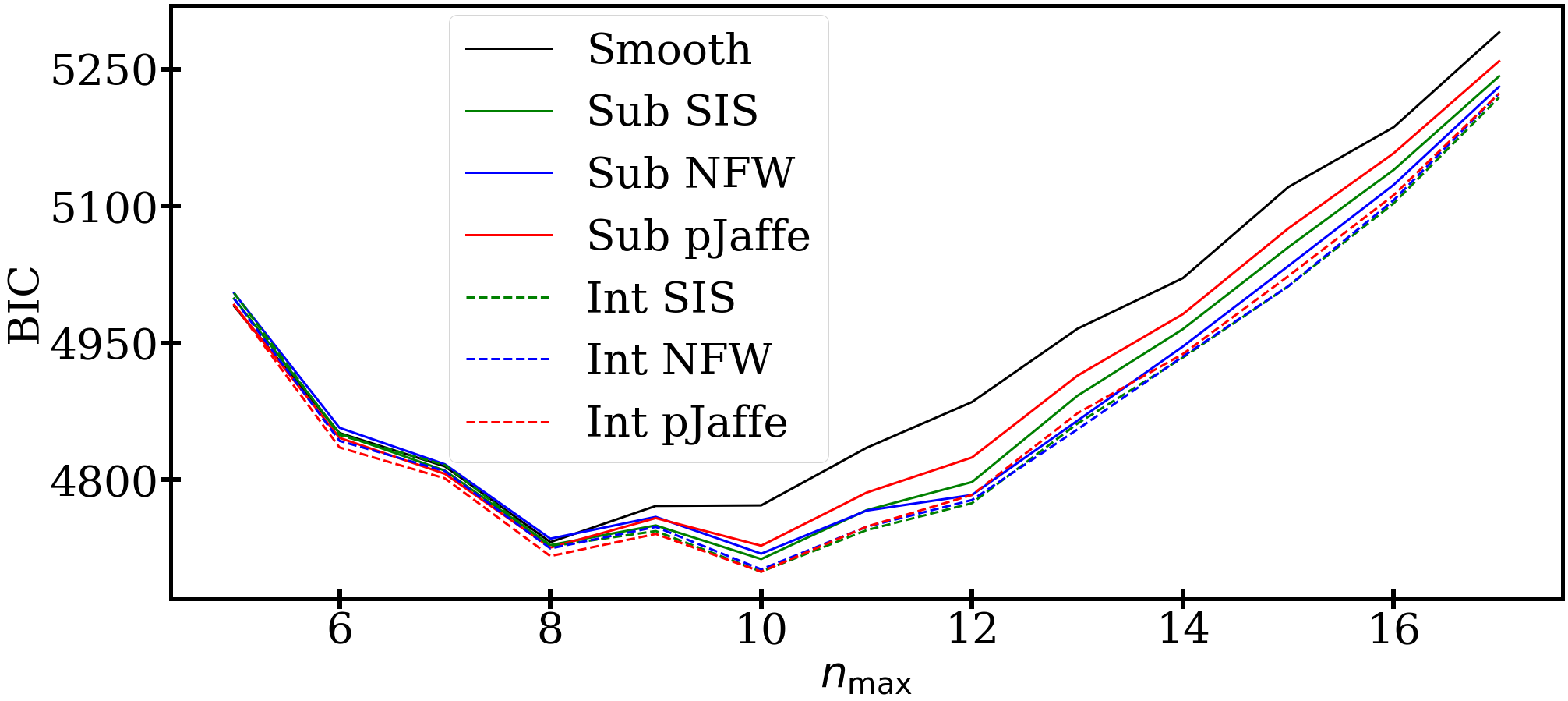}
    \caption{$\mathrm{BIC}$ as a function of the shapelet order parameter $n_\mathrm{max}$ for the different models considered in this work. Shapelets are used to model the source light. We see that $\mathrm{BIC}$ is minimized at $n_\mathrm{max} = 8$ for the smooth model and $n_\mathrm{max} = 10$ for all subhalo and interloper models. Further increasing the source complexity does not continue to improve the fit quality. The degrees of freedom are the number of unmasked pixels subtracted by the number of free parameters.}
    \label{fig:minbic2}
\end{figure}

It is important to be careful about source regularization when modeling gravitational lenses since the source surface brightness has to be reconstructed based on some assumptions about its shape \citep{bayesian_lens_modeling}. We model the source light with shapelets \citep{shapelets,lenstronomy_shape,Birrer:2018vtm,DES:2019fny}, which consist of an orthonormal set of weighted Hermite polynomials. The number of degrees of freedom $N$ in a given shapelet set is determined by the order parameter $n_\mathrm{max}$ as $N = (n_\mathrm{max} + 1)(n_\mathrm{max} + 2)/2$. The scale of the shapelet reconstruction is controlled by the scaling parameter $\delta$. Once all the model parameters are fixed, pixel brightnesses become a linear function of the $N$ shapelet coefficients, whose best fit is analytically solvable. For a fixed $n_\mathrm{max}$, the complexity of the shapelet reconstruction of the source is inherently regularized since it has $N$ degrees of freedom. To choose the ideal $n_\mathrm{max}$ range to correctly regularize the source complexity in the system studied in this work, we iteratively increase $n_\mathrm{max}$ starting from a small value of $n_\mathrm{max}=5$ and run a nested sampling algorithm at each value. For each $n_\mathrm{max}$ we minimize the BIC, where the number of model parameters includes the shapelet coefficients. BIC is the Bayesian information criterion given by $\mathrm{BIC} = k\ln(n) +\chi^2 $,  where $k$ is the number of model parameters and $n$ is the number of data points. Although increasing $n_\mathrm{max}$ always lowers the best fit $\chi^2$, at some point the BIC starts increasing since the number of shapelet degrees of freedom scales with $\propto n_\mathrm{max}^2$. As shown in Fig. \ref{fig:minbic2}, we find the BIC value to be lowest when $n_\mathrm{max} = 8$ for the smooth model and $n_\mathrm{max} = 10$ for all the subhalo and interloper models considered in this work. We will vary $n_\mathrm{max}$ in the neighborhood of this value and show that our results are consistent (see Fig. \ref{fig:nfws} below). By changing $n_\mathrm{max}$, we change the regularization of the source-light modeling. Unless the source-light distribution has brightness fluctuations that are smaller than the smallest shapelet scale, which decreases as $n_\mathrm{max}$ increases, we expect consistency over a small range of $n_\mathrm{max}$ values. 

In addition to checking that our results are consistent when varying the source-light regularization, we also perform robustness checks on
the mask.
To test if masking introduces any systematic changes, we tried changing the width of the unmasked annulus by 30\% and also got consistent results. 

\subsection{Nested Sampling}
\begin{figure}
    \centering
    \includegraphics[width=\linewidth]{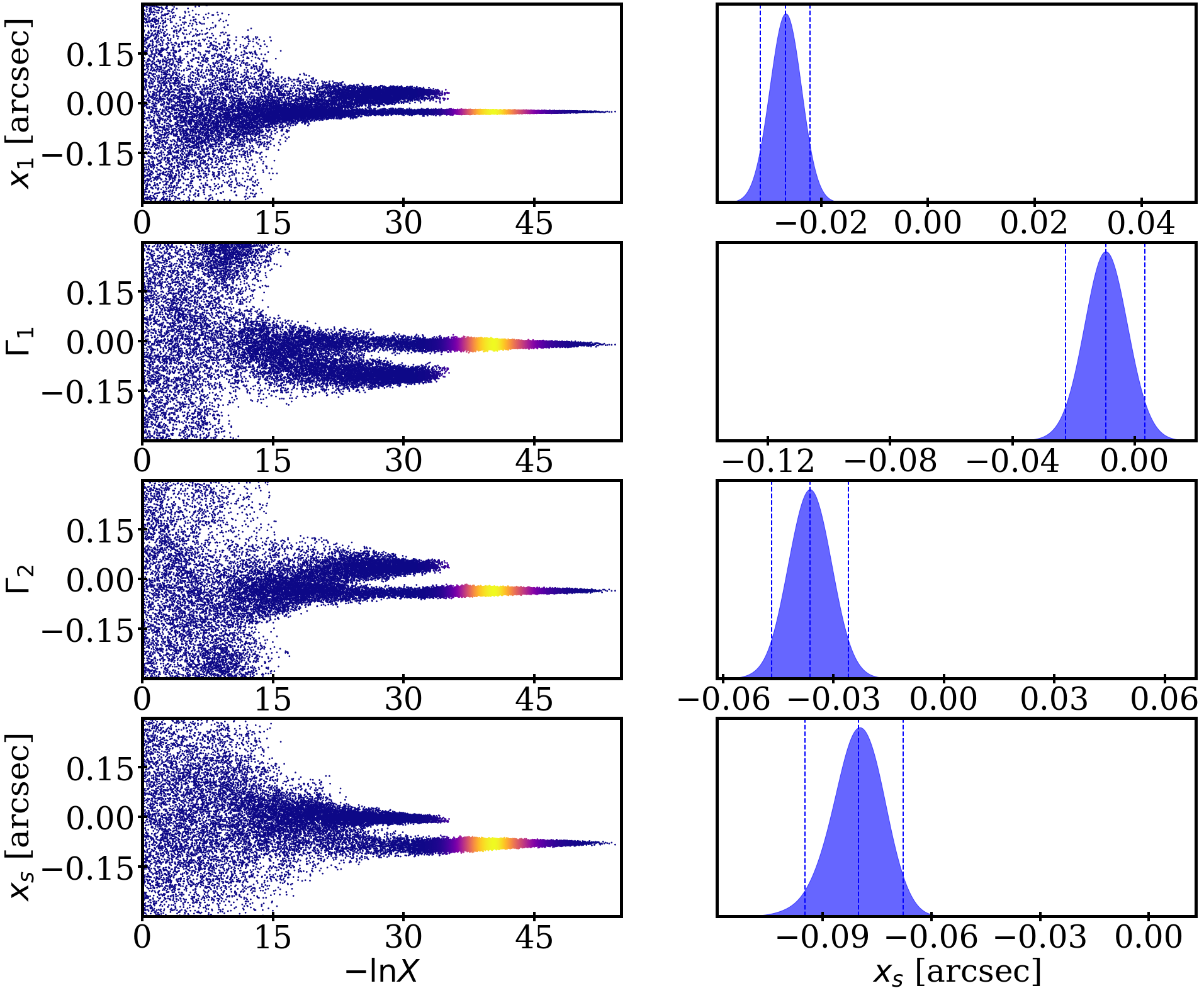}
    \caption{\textit{Left: }Trace plots of some of the lensing parameters for the pseudo-Jaffe subhalo model ($M_\mathrm{subpJaffe10}$ in Table \ref{table:1}) on real data with $n_\mathrm{max}=10$, taken as an example to show the local maxima of the likelihood. The parameter $x_1$ corresponds to the center of the main lens potential, $\Gamma_1$ and $\Gamma_2$ are the components of the external shear, and $x_s$ is the $x$ coordinate of the center of the source-light distribution. $X$ is the fractional prior volume that has a higher likelihood than the sampled point. Points with smaller $X$ (higher -$\ln X$) correspond to points that have a higher likelihood. \textit{Right: } Marginalized posterior probability distributions of the lensing parameters for the pseudo-Jaffe subhalo model ($M_\mathrm{subpJaffe10}$ in Table \ref{table:1}) on real data with $n_\mathrm{max}=10$.}
    \label{fig:multi_peak}
\end{figure}

The posteriors of lensing parameters include many local maxima (see Fig. \ref{fig:multi_peak} for an example), which makes finding best fits and sampling the likelihood with Markov chains challenging. 
Therefore, we apply a dynamic nested sampling algorithm to obtain our best fits and posteriors. We use a publicly available package known as \texttt{dynesty} \citep{2020MNRAS.493.3132S}.

\subsection{Mock Images and Modeling Biases}

\begin{figure}
    \centering
    \includegraphics[width=\linewidth]{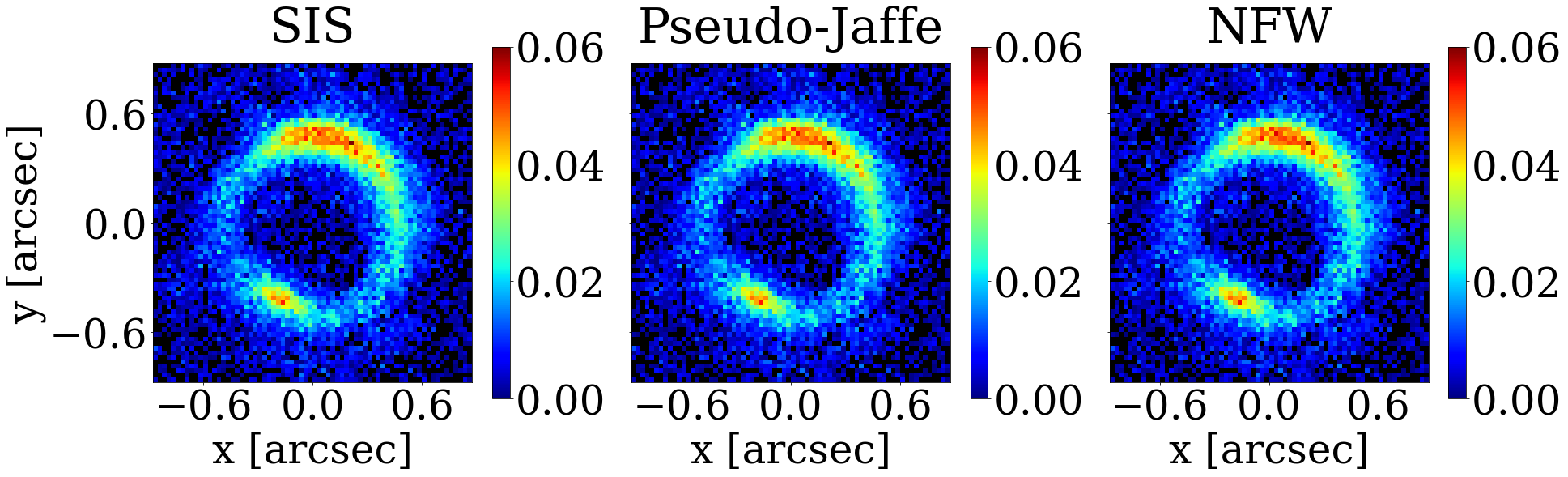}
    \caption{70x70 pixels, 1.6 $\mu$m mock images with interlopers of different mass profiles.}
    \label{fig:mock_fg}
\end{figure}

In order to test for the feasibility of measuring the redshift of the perturber and to investigate any biases that the different models can introduce, we created  mock images resembling the JVAS B1938+666 system (see Fig. \ref{fig:mock_fg}). The background and Poisson noise of each mock image are set to match the noise levels of the HST image used in this study. The former is due to the electronic noise in the detector and badly masked light sources near the image and the latter is due to the statistical fluctuation in the number of photons that are received from the source of interest. The source is created with a basis of shapelets with $n_\mathrm{max} = 10$ 
and a value of $\delta$ obtained from the corresponding best fit on real data for each model. The three mock images have an interloper at $z_\mathrm{int}=1.427,\,1.417,\,1.420$, with an SIS, pseudo-Jaffe and NFW profile, respectively. Each image is then analyzed with a model that includes its own true profile, and the two other profiles. We found that the analysis of the parameters of the perturber is highly model dependent. As expected, the true model gives the best fit of the data in each case. Using a different model for the perturber still gives a statistically significant improvement over a smooth model. However, it can introduce biases to the inferred mass of the perturber. Notably, the inferred redshift is robust under these changes.

\begin{figure}
    \centering
    \includegraphics[width=\linewidth]{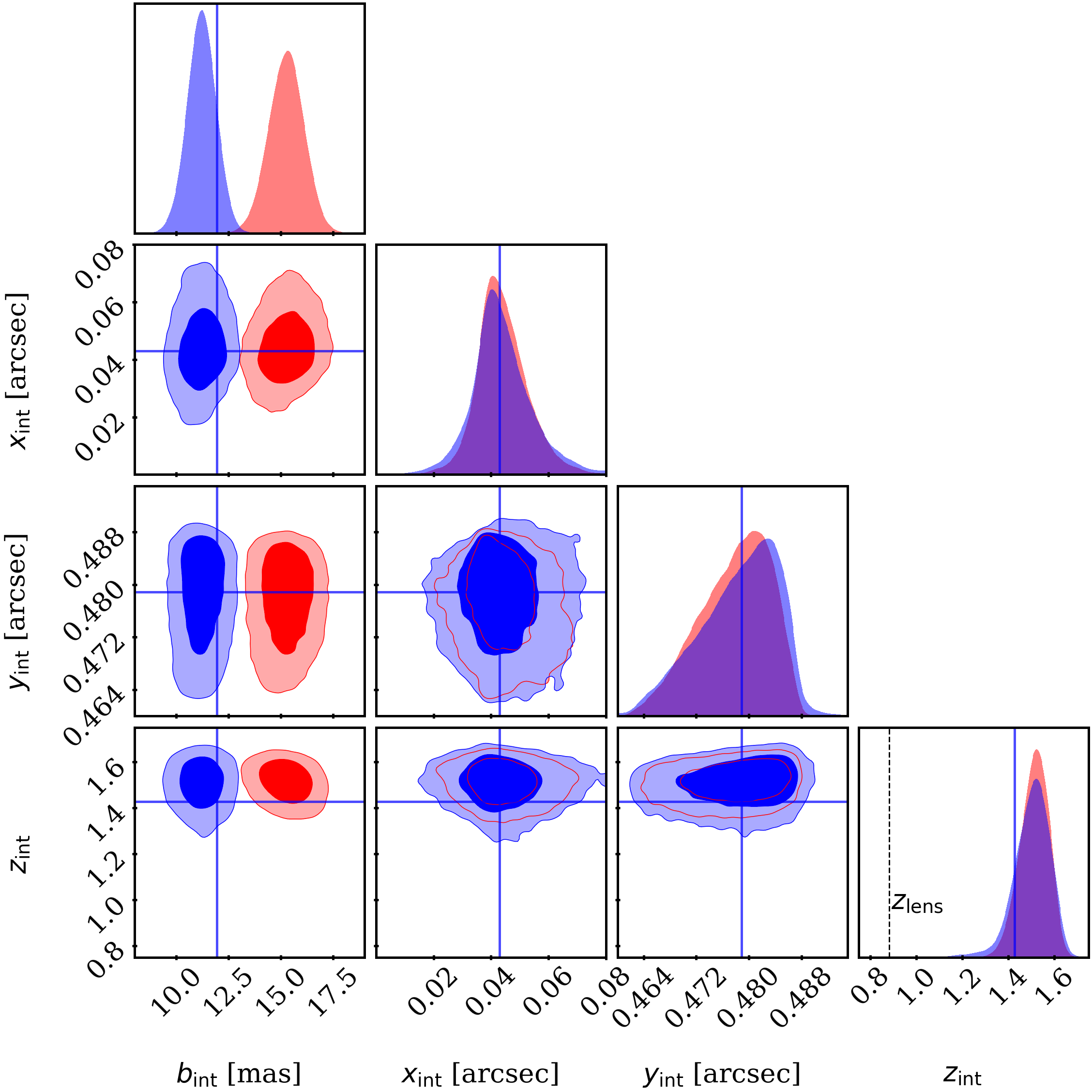}
    \caption{Posterior probability distributions of the interloper parameters of pseudo-Jaffe (red) and SIS (blue) on a \textit{mock} image created with an SIS, shown in Fig. \ref{fig:mock_fg}. The true values are shown as solid lines.}
    \label{fig:mockbshift}
\end{figure}

\begin{figure}
    \centering
    \includegraphics[width=\linewidth]{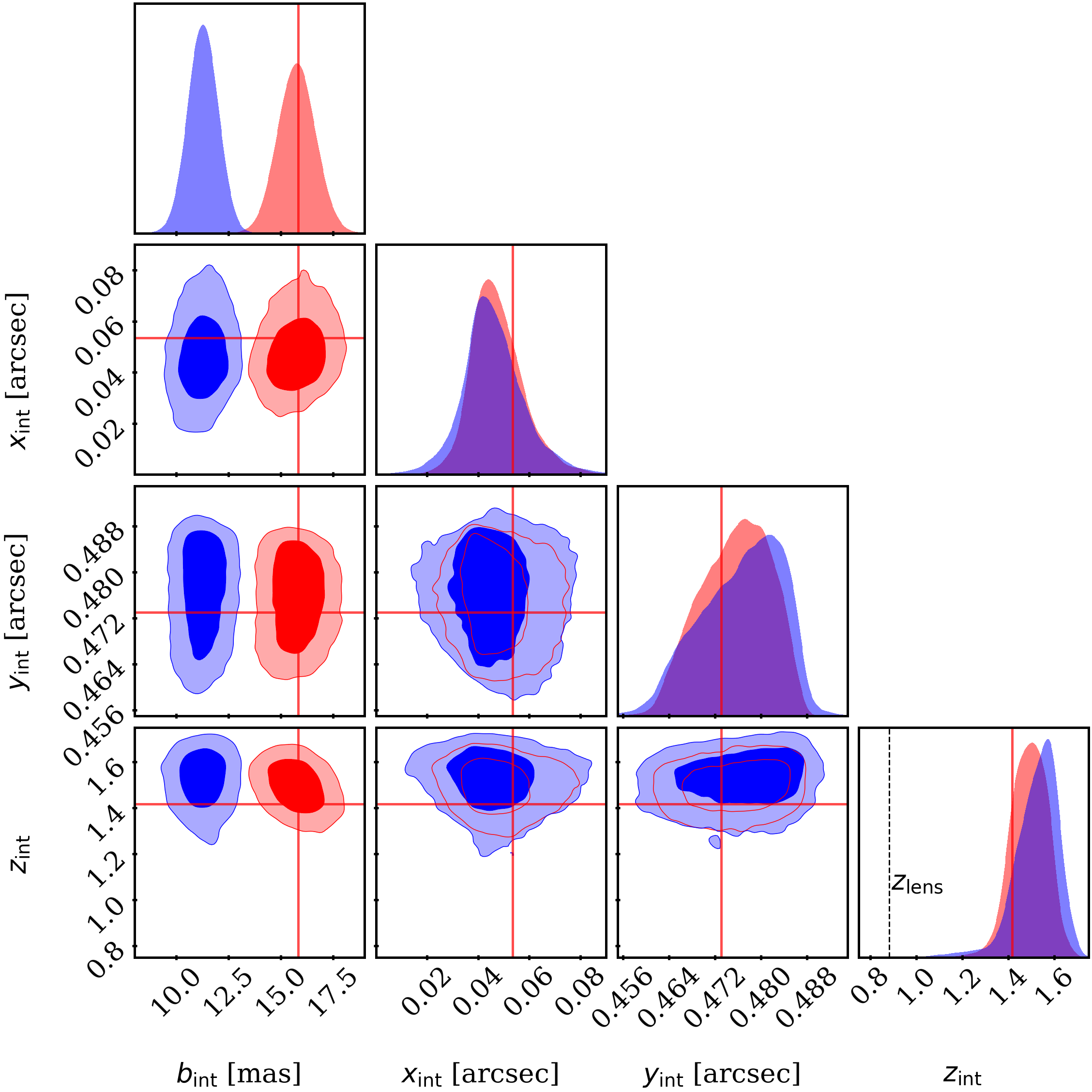}
    \caption{Posterior probability distributions of the interloper parameters of pseudo-Jaffe (red) and SIS (blue) on a \textit{mock} image created with a pseudo-Jaffe, shown in Fig. \ref{fig:mock_fg}. The true values are shown as solid lines.}
    \label{fig:mockbshift2}
\end{figure}

The difference between an NFW and SIS is clear since the 3D density of the former scales as $r^{-1}$ inside the scale radius and $r^{-3}$ outside, while the latter scales as $r^{-2}$ everywhere. This creates different magnifications and deflection angles near the perturber where the effects are strongest. The difference between SIS and pseudo-Jaffe is more subtle. In the $r\rightarrow 0$ limit,
\begin{equation}\label{eq:diflims}
    \kappa_\mathrm{pJaffe}(r) \cong  \frac{b_\mathrm{int}}{2}\left[\frac{1}{r} - \frac{1}{r_t}\right] = \kappa_\mathrm{SIS}(r) - \frac{b_\mathrm{int}}{2r_t} .
\end{equation}
Thus, pseudo-Jaffe includes an SIS term minus a constant convergence. To counter this offset, the best fit $b_\mathrm{int}$ for a pseudo-Jaffe is larger than that of SIS (see Figs. \ref{fig:mockbshift} and \ref{fig:mockbshift2}). This implies that obtaining a $b_\mathrm{int}$ for a perturber from a model with an SIS and inferring its total mass from that value by assuming it is truncated will give a value lower than the true mass of the perturber.

\begin{figure}
    \centering
    \includegraphics[width=\linewidth]{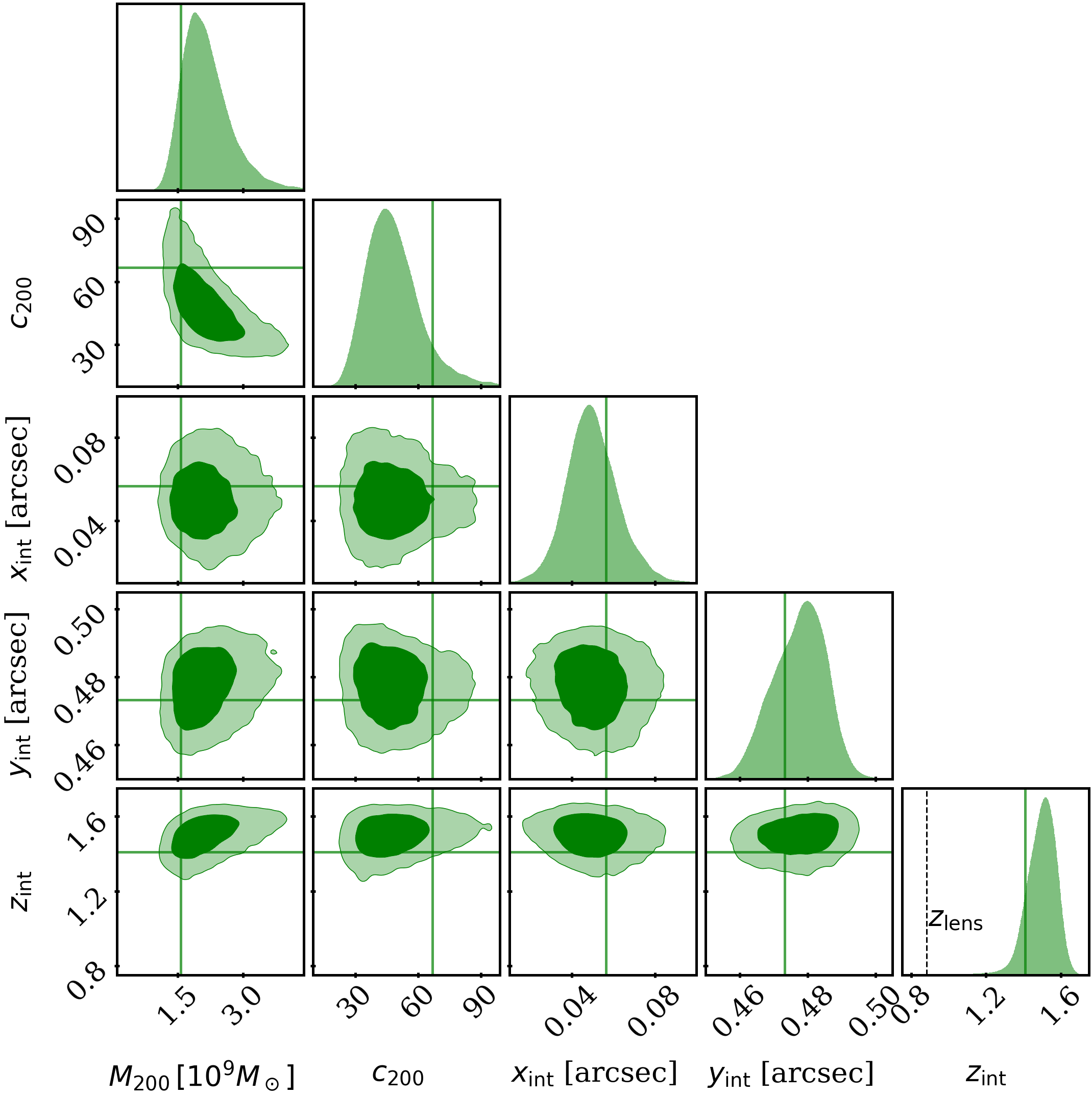}
    \caption{Posterior probability distributions of the interloper parameters of NFW on a \textit{mock} image created with an NFW, shown in Fig. \ref{fig:mock_fg}. The true values are shown as solid lines.}
    \label{fig:mocknfw}
\end{figure}

\begin{figure}
    \centering
    \includegraphics[width=\linewidth]{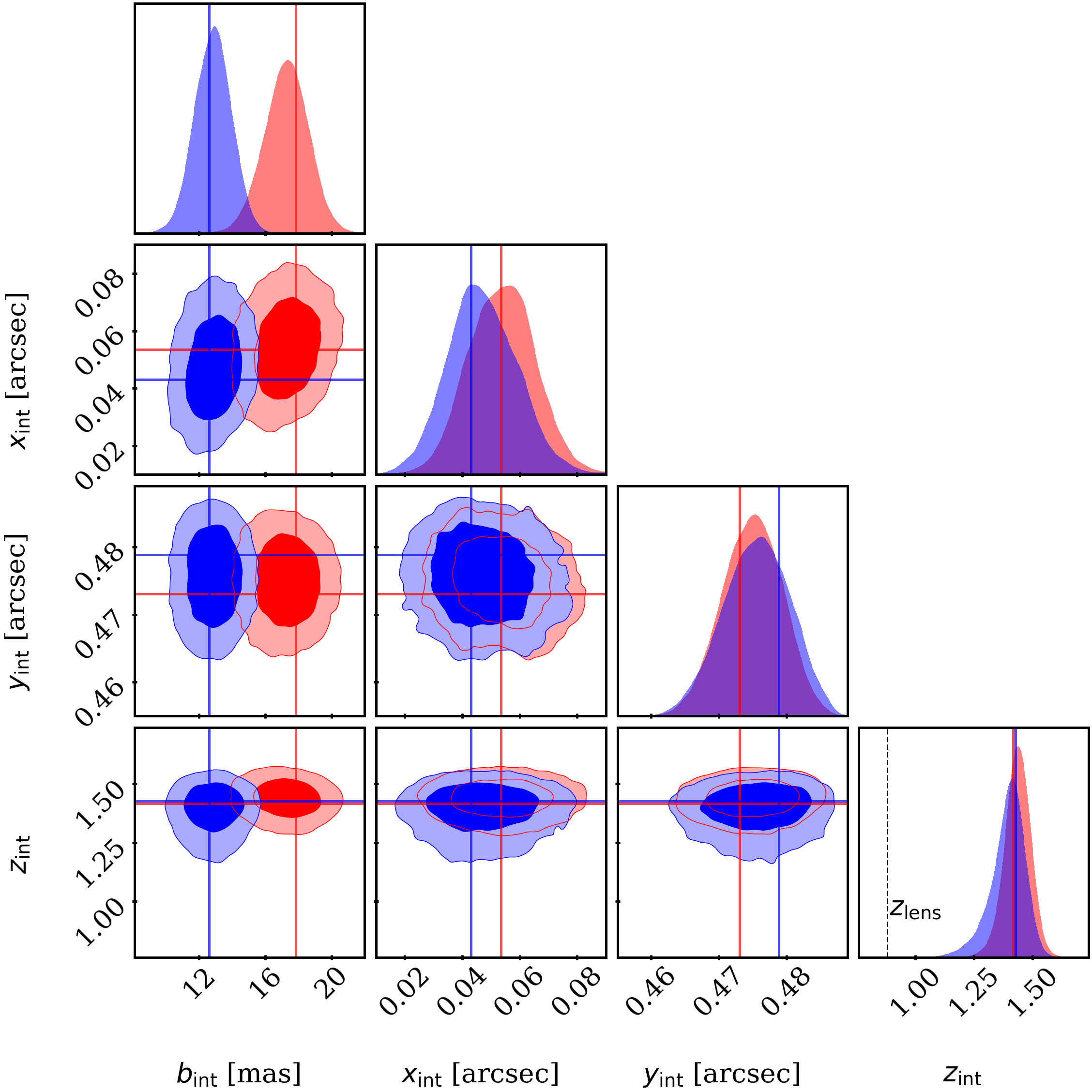}
    \caption{Posterior probability distributions of the interloper parameters modeling the real data with $M_\mathrm{intSIS10}$ (blue) and $M_\mathrm{intpJaffe10}$ (red), described in Table \ref{table:1}. The best fits are shown as solid lines.}
    \label{fig:sisjaf}
\end{figure}

\begin{figure}
    \centering
    \includegraphics[width=\linewidth]{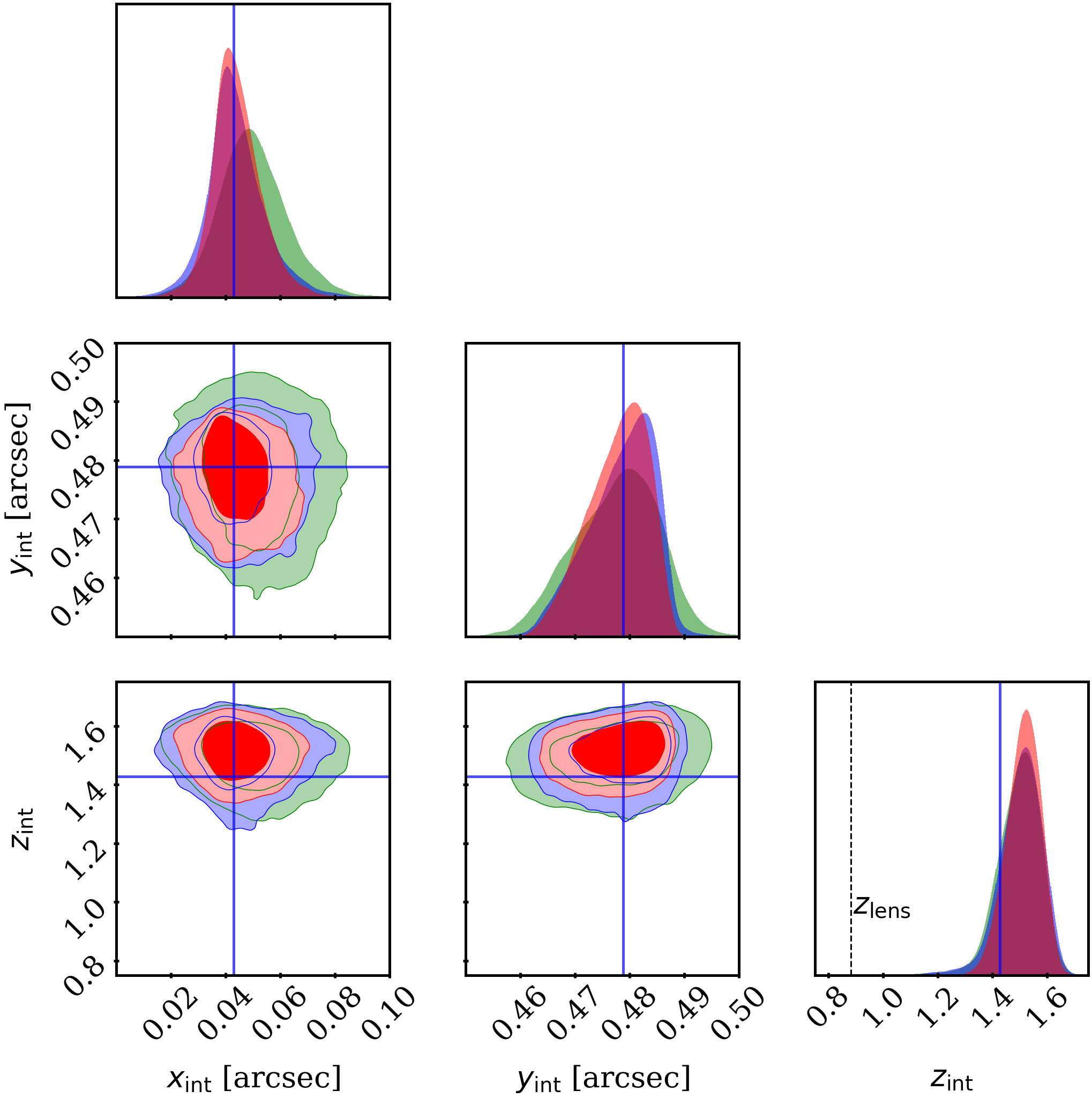}
    \caption{Posterior probability distributions of the interloper parameters of pseudo-Jaffe (red), SIS (blue),  and NFW (green) on a \textit{mock} image created with an SIS shown in Fig. \ref{fig:mock_fg}. The true values are shown as solid lines.}
    \label{fig:zchange}
\end{figure}

In gravitational lensing, there is a mass-redshift correlation due to the relevant quantity that does the lensing being the convergence, not the mass density. Mass is weighted by the critical surface density $\Sigma^{-1}_\mathrm{c}$, which can be seen in Eq. \eqref{eq:profiles2}. The critical surface density is a function of the redshift of the lens and the source. The resulting mass-redshift correlation can be seen from the analysis of NFW mocks in Fig. \ref{fig:mocknfw} and in the real data (Fig. \ref{fig:mass_concen}). This degeneracy is not seen in the posteriors of models with pseudo-Jaffe and SIS models (as can be seen in Figs. \ref{fig:mockbshift}, \ref{fig:mockbshift2}, and \ref{fig:sisjaf}) because we use $b_\mathrm{int}$ as the model parameter, which directly controls the amount of convergence. In Fig. \ref{fig:zchange} we compare the three models analyzed (NFW, SIS and pseudo-Jaffe) and find that the parameters for the position and the redshift of the perturber are consistent (with NFW giving generally larger uncertainties).

We assume flat priors for the apparent position as well as the redshift of the interloper. This corresponds to non-flat priors at the plane of the interloper for background interlopers due to lensing by the main lens. A flat redshift prior overestimates the prior probability of the interloper being at higher redshift since comoving distance increases more slowly with increasing redshift. Additionally, a flat redshift prior also overestimates the prior probability of the interloper being further from the main lens, because the comoving area of the cross section of the line-of-sight volume is not uniform (it is widest at the main lens). However, we find that our choice of flat priors have a negligible effect on the posteriors of the model parameters.

\begin{figure}
    \centering
    \includegraphics[width=1.\linewidth]{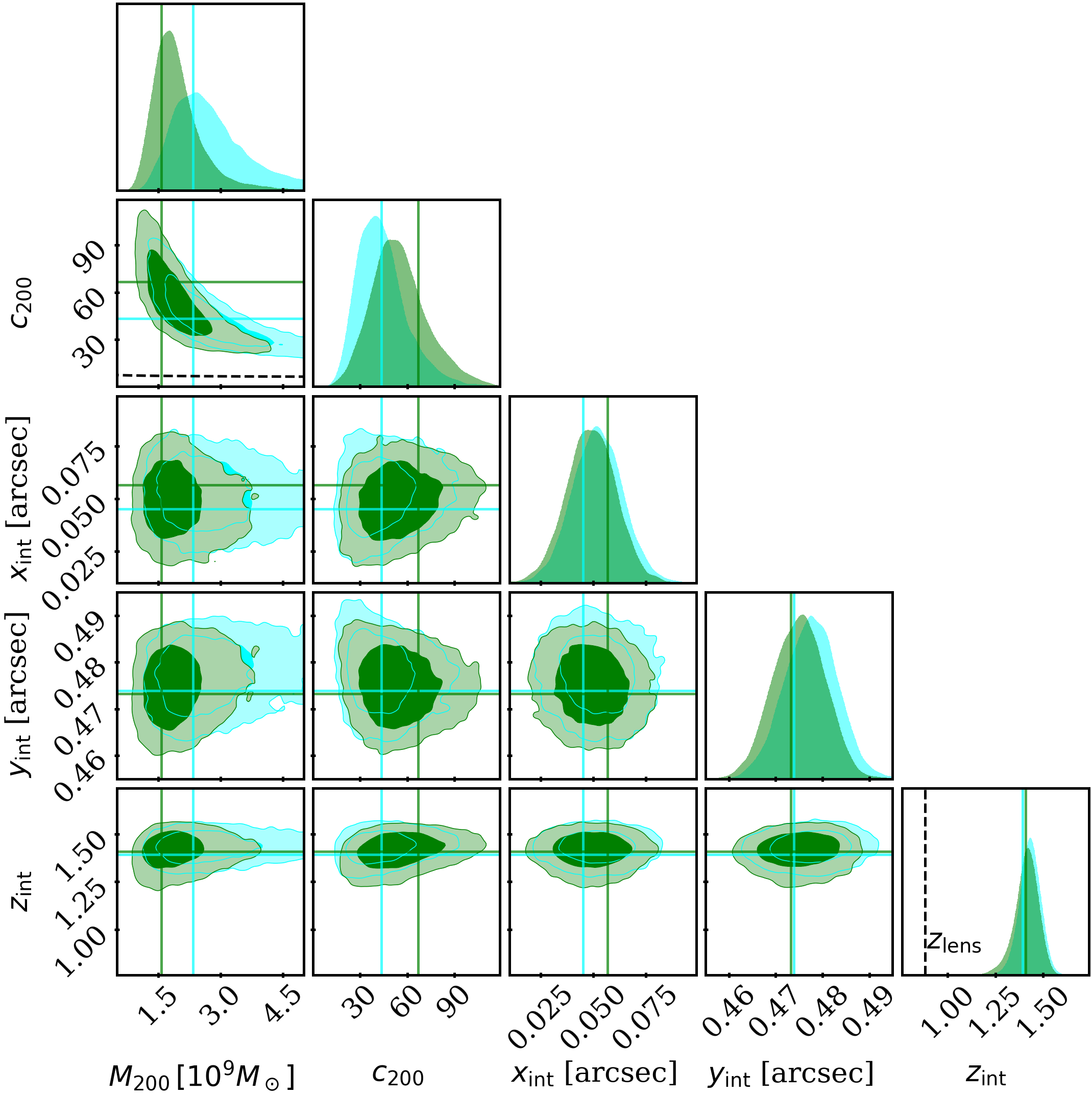}
    \caption{Posterior probability distributions of the interloper parameters in the real data modeled as an NFW profile with the source light modeled by a shapelet set with order parameters $n_\mathrm{max} = 10$ and $11$ shown in green and cyan, respectively. The best fits are shown as solid lines. The dashed horizontal line corresponds to the mass-concentration relation at the best fit redshift of the interloper \citep{mass_concen}. The vertical dashed line corresponds to the redshift of the main lens.}
    \label{fig:nfws}
\end{figure}

\begin{figure}
    \centering
    \includegraphics[width=\linewidth]{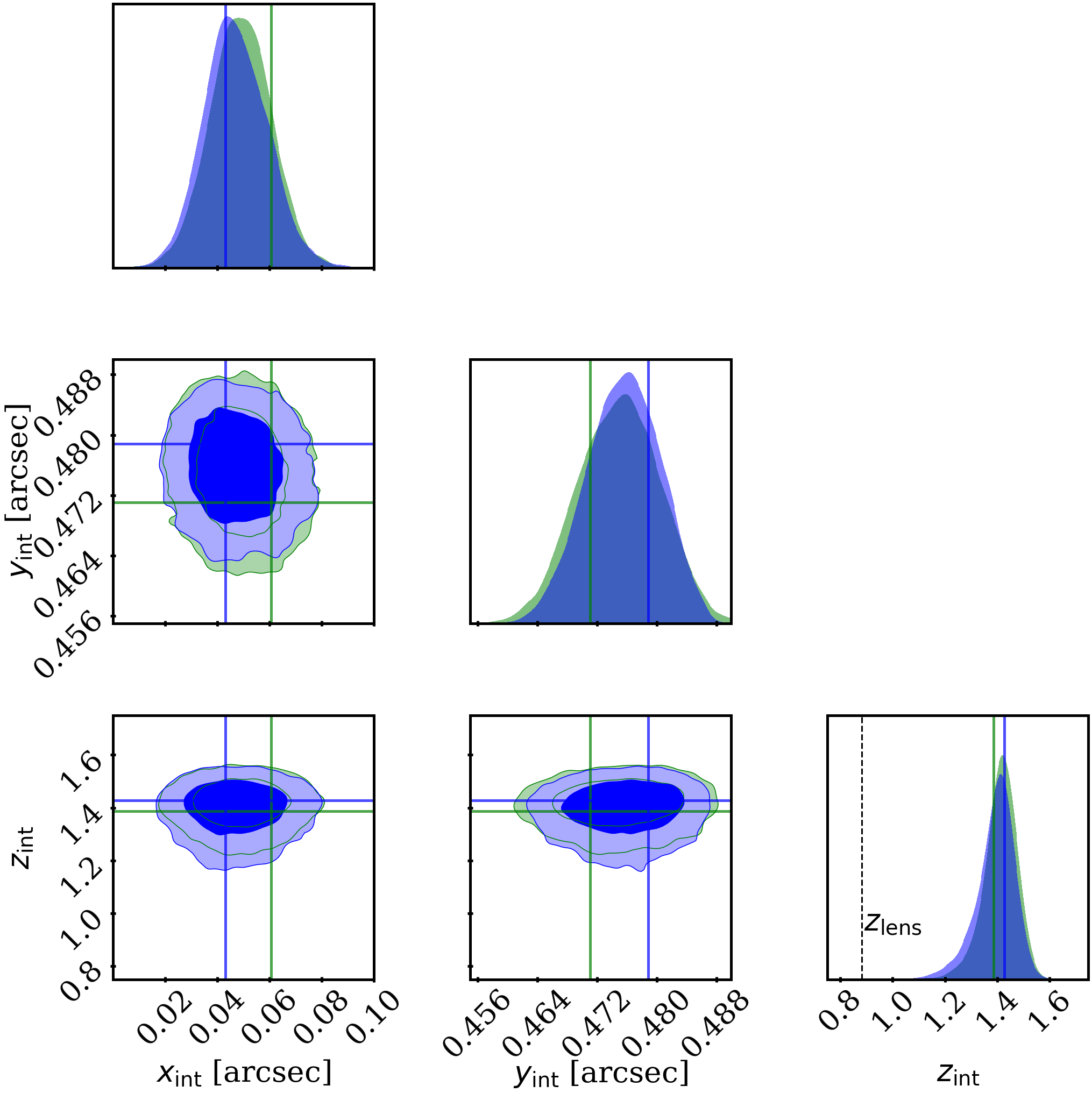}
    \caption{Posterior probability distributions of the interloper parameters modeling the real data with $M_\mathrm{intSIS10}$ (blue) and $M_\mathrm{intNFW10}$ (green), described in Table \ref{table:1}. The best fits are shown as solid lines.}
    \label{fig:sisnfw}
\end{figure}

\begin{table*}
\centering
\begin{tabular}{|p{4.5cm}||p{2.5cm}|p{2.5cm}|p{4.5cm}|}
\hline
Model Name&$n_\mathrm{max}$&Type&Perturber Mass Profile\\
\hline
$M_\mathrm{smooth8}$&8&Smooth&--\\
$M_\mathrm{smooth10}$&10&Smooth&--\\
$M_\mathrm{subNFW10}$&10&Subhalo&NFW\\
$M_\mathrm{intNFW10}$&10&Interloper&NFW\\
$M_\mathrm{subSIS10}$&10&Subhalo&SIS\\
$M_\mathrm{intSIS10}$&10&Interloper&SIS\\
$M_\mathrm{subpJaffe10}$&10&Subhalo&pseudo-Jaffe\\
$M_\mathrm{intpJaffe10}$&10&Interloper&pseudo-Jaffe\\

\hline

\end{tabular}

\caption{Descriptions of the different models used to fit the image. In general, our notation will be $M_\mathrm{model n_{\rm max}}$.}
\label{table:1}
\end{table*}

\subsection{Real Data}
The analysis of the real data uses the same pipeline as the mock images discussed earlier. The data (that had been previously \textit{drizzled}) is obtained from the Hubble Legacy Archive. We analyze the system with $8$ different models that are listed in Table \ref{table:1}.  The shapelet order parameter $n_\mathrm{max}$ is set to be either $n_\mathrm{max} = 8$ or $n_\mathrm{max} = 10$,  determined by the lowest $\mathrm{BIC}$ that we get from smooth and subhalo/interloper models,  respectively, shown in Fig. \ref{fig:minbic2}. We show in Table \ref{table:2} how the fit improves when an NFW subhalo is added to a smooth model and when the subhalo is changed to an interloper. The model with an NFW subhalo is preferred over a smooth model with a $\Delta(\mathrm{BIC}) = -12.6$. We find a further improvement of $\Delta(\mathrm{BIC}) = -17.2$ for a model with an NFW interloper over a model with an NFW subhalo, and a Bayes factor of $\log_{10}{\rm K}=3.4$. Table \ref{table:2} also shows a similar improvement when using SIS. The improvement for a pseudo-Jaffe interloper over a subhalo is greater than the other two cases, but this is only because the improvement from adding a pseudo-Jaffe subhalo to a smooth model is smaller.
The best fits for the lensing strength $b_\mathrm{int}$ of the pseudo-Jaffe model are larger than for SIS due to the different $r\rightarrow 0$ limits that the profiles have, as shown in Eq. \eqref{eq:diflims}.

\begin{table*}
\centering
\begin{tabular}{|p{2.0cm}||p{1.4cm}|p{1.4cm}|p{1.4cm}|p{1.4cm}|p{1.4cm}|p{1.4cm}|p{1.4cm}|p{1.4cm}|}
\hline
Parameter&$M_\mathrm{smooth8}$&$M_\mathrm{smooth10}$&$M_\mathrm{subSIS10}$&$M_\mathrm{intSIS10}$&$M_\mathrm{subNFW10}$&$M_\mathrm{intNFW10}$&$M_\mathrm{subpJaffe10}$&$M_\mathrm{intpJaffe10}$\\
\hline
$\theta_E$&0.467$''$&0.467$''$&0.457$''$&0.460$''$&0.459$''$&0.463$''$&0.465$''$&0.464$''$\\
$\gamma$&2.530&2.470&2.340&2.313&2.320&2.322&2.357&2.348\\
$x_1$&-0.037$''$&-0.031$''$&-0.025$''$&-0.025$''$&-0.023$''$&-0.025$''$&-0.027$''$&-0.024$''$\\
$x_2$&-0.111$''$&-0.108$''$&-0.111$''$&-0.108$''$&-0.110$''$&-0.108$''$&-0.107$''$&-0.108$''$\\
$e_1$&-0.020&-0.080&-0.020&-0.001&-0.011&0.004&-0.038&-0.010\\
$e_2$&0.110&0.104&0.064&0.059&0.064&0.056&0.076&0.065\\
$\Gamma_1$&0.010&-0.010&-0.009&0.003&-0.011&0.004&-0.009&-0.001\\
$\Gamma_2$&-0.043&-0.038&-0.035&-0.034&-0.034&-0.035&-0.032&-0.037\\
\hline
$b_\mathrm{int}(\times 10^{-3})$&--&--&12.8$''$&12.6$''$&--&--&17.2$''$&17.8$''$\\
$M_{200}/M_\odot (\times 10^{9})$&--&--&--&--&1.96-&1.56&--&--\\
$c_{200}$&--&--&--&--&23.47&66.75&--&--\\
$x_\mathrm{int}$&--&--&0.036$''$&0.043$''$&0.035$''$&0.057$''$&0.037$''$&0.054$''$\\
$y_\mathrm{int}$&--&--&0.473$''$&0.479$''$&0.480$''$&0.473$''$&0.480$''$&0.473$''$\\
\hline
$z_\mathrm{int}$&--&--&0.881(fixed)&1.43&0.881(fixed)&1.42&0.881(fixed)&1.42\\

\hline
BIC&4731.88&4772.14&4713.31&4699.55&4719.24&4702.03&4727.94&4699.45\\
\hline
 $\log_{10} \epsilon$ &-943.35&-914.06&-899.53&-896.30&-899.46&-896.04&-902.14&-896.36\\
\hline

\end{tabular}               
\caption{Real data best fit values for smooth and perturber models, comparing different interloper profiles, along with the BIC values for each model and the Bayesian evidence $\epsilon$. The fit improves for all models considered when adding a subhalo and further improves when replacing it with an interloper. The difference between interloper fits are relatively small compared to the improvement over a subhalo fit. Although it is not our best fit smooth model, we include here $M_\mathrm{smooth10}$ as a comparison with the other models of the same $n_\mathrm{max}$ value. All priors used were non-informative.}
\label{table:2}
\end{table*}

We compare the posterior probabilities of the interloper parameters of $M_\mathrm{intNFW10}$ and $M_\mathrm{intNFW11}$ in Fig. \ref{fig:nfws}. We find that the two models give consistent results for the mass, position, and redshift of the interloper. We see the same mass-redshift degeneracy that was observed in mock data, shown in Fig. \ref{fig:mocknfw}. The mass, concentration, and position uncertainties for the $n_\mathrm{max}=11$ model are slightly larger because a higher shapelet index allows the source light to have more degrees of freedom. Remarkably, the redshift uncertainty is robust under the change of $n_{\rm max}$. It is also worth noting that the concentration parameter is higher than what one would expect from the mass-concentration relation at the redshift of the interloper (see horizontal dashed line), showing a similar trend to what has been observed in other systems \citep{Minor:2020hic}. Fig. \ref{fig:sisnfw} compares the position and the redshift of $M_\mathrm{intSIS10}$ and $M_\mathrm{intNFW10}$ (the masses are not directly comparable). These models give consistent uncertainties in the position estimate of the interloper. We also observe this consistency in our mock data analysis shown in Fig. \ref{fig:zchange}. Fig. \ref{fig:sisjaf} compares the posteriors of $M_\mathrm{intSIS10}$ and $M_\mathrm{intpJaffe10}$. We see that the lensing strength $b_\mathrm{int}$ is significantly higher for the model with a pseudo-Jaffe over that with an SIS (as explained before), similar to what we have seen in our study of the mock images, shown in Fig. \ref{fig:mockbshift} and \ref{fig:mockbshift2}. 

\section{Statistical Interpretation of Future Detections}
We calculate the implications of distinguishing interlopers from subhalos in the context of constraining the fractional mass in substructure, $f_\mathrm{sub}$, and the subhalo mass function slope, $\beta$.  We also discuss the prospects of probing dark matter using interlopers by measuring the half-mode mass $M_\mathrm{hm}$, which is defined as the mass scale at which the warm dark matter (WDM) power spectrum is one half of the CDM power spectrum. The ultimate goal is to use these observables to distinguish between different dark matter scenarios.

\subsection{Subhalo Mass Function Normalization}
To illustrate how the constraints on the subhalo mass function can be affected if we count interlopers as subhalos, we will consider a scenario where we are able to detect all perturbers with mass above $m_\mathrm{eff}\geq 10^{8} \mathrm{M}_\odot$. We also assume that the perturbers are within a $0.1''$ wide annulus around the Einstein radius, which typically includes the brightest source-light pixels. Previous studies, which take into account the changes in the lowest detectable masses in different parts of the image due to different pixel brightnesses, have constrained dark matter properties using the lack of subhalo or line-of-sight halo detections in the known strong-lensing systems \citep{lowest_detectable_mass,lack_of_detection_constraint}.  We divide the perturbers into 30 logarithmic mass bins between $10^{8}-10^{11}\mathrm{M}_\odot$. We pick JVAS B1938+666 as a reference system, which gives us an annulus with a physical area $A\approx18\,\mathrm{kpc}^2$. The number of expected subhalos within this area in each mass bin is then given as
\begin{equation}
    \mu_{sub,i} = A \int^{M_{i,2}}_{M_{i,1}} dM\, n_\mathrm{sub}(M),
\end{equation}
where $n_\mathrm{sub} \propto M^{\beta}$ is the subhalo mass function per unit area normalized such that the total mass of the subhalos is $f_\mathrm{sub}$ times the mass of the host. $M_{i,1}$ and $M_{i,2}$ are the lower and upper bounds of the mass bin $i$. If $n_{sub,i}$ is the number of subhalos detected in mass bin $i$, the log-likelihood, up to a normalization, is given by \citep{poisson_chi}
\begin{equation}\label{eq:sub_poisson}
    \ln \lambda(f_\mathrm{sub},\beta) = -\sum^{N=30}_{i=1}\left[ \mu_{\mathrm{sub},i} + \ln \frac{ n_{\mathrm{sub},i}!}{[\mu_{\mathrm{sub},i}]^{n_{\mathrm{sub},i}}}\right],
\end{equation}
where the last term is zero if $n_{\mathrm{sub},i}=0$. We generate a random Poissonian realization $n_{\mathrm{sub},i}$ to model the number of detections at each mass bin with the subhalo mass function parameters set to $f_\mathrm{sub} = 0.03$, $\beta = -1.9$. We then use Eq. \eqref{eq:sub_poisson} to forecast the uncertainties in $f_\mathrm{sub}$ and $\beta$.
We calculate the expected number of interlopers for the same effective area. Interlopers will populate a double-cone shaped volume whose cross section depends on the interloper, lens, and source redshift, and whose base is the effective area $A$ that we have selected earlier. We have
\begin{align}\label{eq:cone_area}
    \frac{S(\chi)}{A/a^2(\chi_l)} = \left\{
        \begin{array}{ll}
            \left(\dfrac{\chi}{\chi_l}\right)^2 & \quad \chi \leq \chi_l \\[3ex]
            \left(\dfrac{\chi_s - \chi}{\chi_s - \chi_l}\right)^2 & \quad \chi > \chi_l
        \end{array},
    \right.
\end{align}
where $\chi,\chi_l$, and $\chi_s$ are the comoving distances to the interloper, main lens, and source, $S(\chi)$ is the comoving area of the cross section, and $a(\chi_l)$ is the scale factor at the lens \citep{int_pow_spec}. So the expected number of interlopers at each mass bin $i$ is given by
\begin{equation}
    \mu_{\mathrm{int},i} = \int^{\chi_s}_{0}d\chi \int^{m_\mathrm{eff}(M_{i,2},\chi)}_{m_\mathrm{eff}(M_{i,1},\chi)}dM\, n_\mathrm{int}(M) S(\chi),
\end{equation}
where $m_\mathrm{eff}(M,\chi)$ is the effective mass function that maps the true mass of the interloper to its effective mass as if it were a subhalo on the lens plane \citep{int_pow_spec}. 

\begin{figure}
    \centering
    \includegraphics[width=0.7\linewidth]{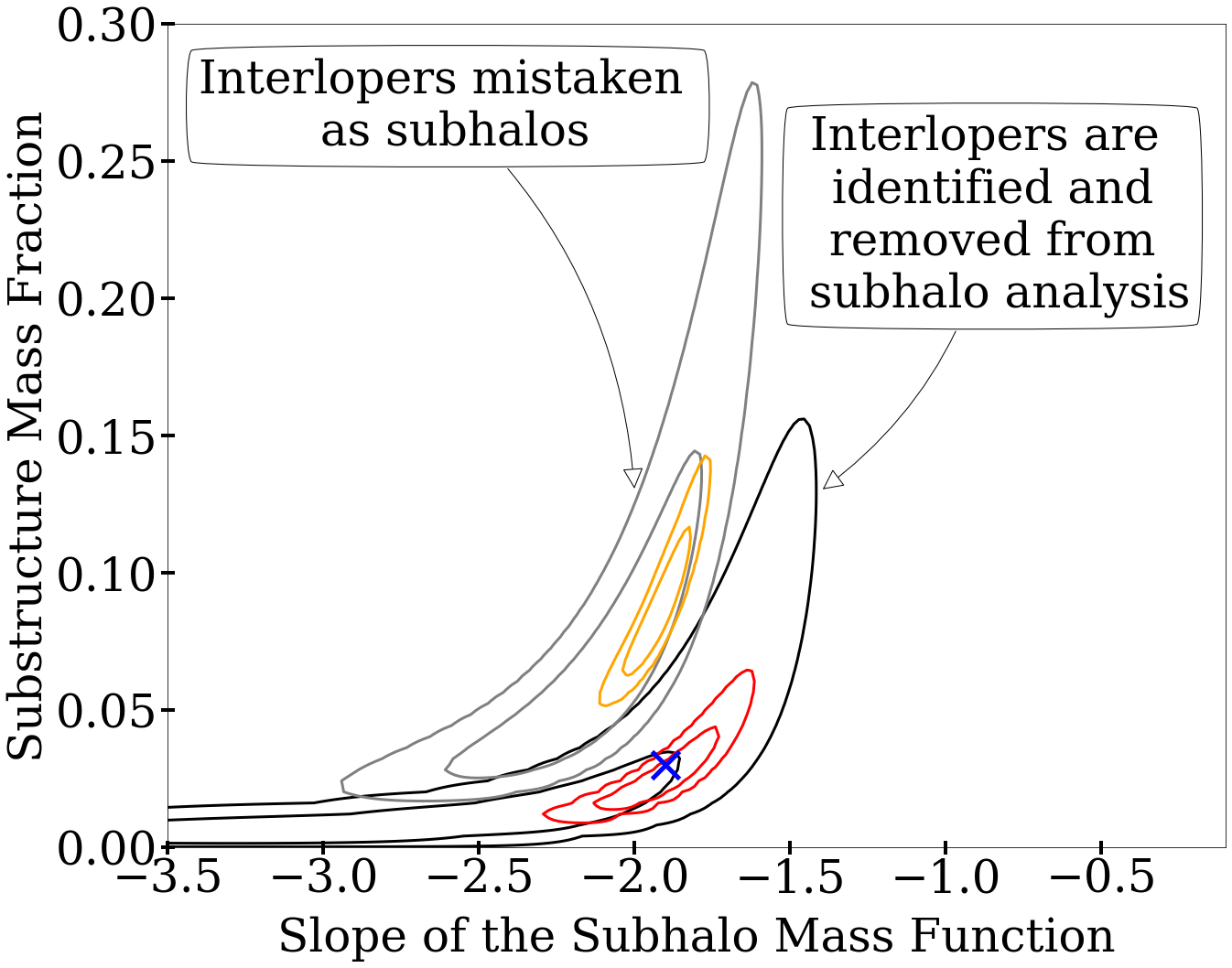}
    \caption{Correctly identifying line-of-sight halos is crucial for analyzing the subhalo mass function. Shown here are the 1-2 $\sigma$ contours of the two subhalo mass function parameters for mock interloper and subhalo populations.  The true value is shown as the blue cross. The black (red) contours show the case when we can correctly identify the interlopers and remove them from the analysis of the subhalo population when we probe 10 (90) times the effective area of JVAS B1938+666. In contrast, the grey and orange contours count all the perturbers within the volume as subhalos for 10 and 90 times the effective area, respectively.}    
    \label{fig:stat_fig1}
\end{figure}

We also generate random Poissonian realizations $n_\mathrm{int,i}$ to model the number of interloper detections at each mass bin. We show how the constraints on $f_\mathrm{sub}$ and $\beta$ are affected if we count these interlopers as subhalos. In Fig. \ref{fig:stat_fig1}, we show what happens when the interlopers are confused as subhalos in the cases of $21$ and $176$ total perturber detections. These numbers are the expected number of detections if we are able to probe an effective angular area in the sky that corresponds to 10 and 90 times the area of our reference system, respectively. As one might expect, it results in an overestimate for the $f_\mathrm{sub}$ which affects the constraints on dark matter that use the substructure mass function.

\subsection{Half-mode Mass}

\begin{figure}
    \centering
    \includegraphics[width=0.7\linewidth]{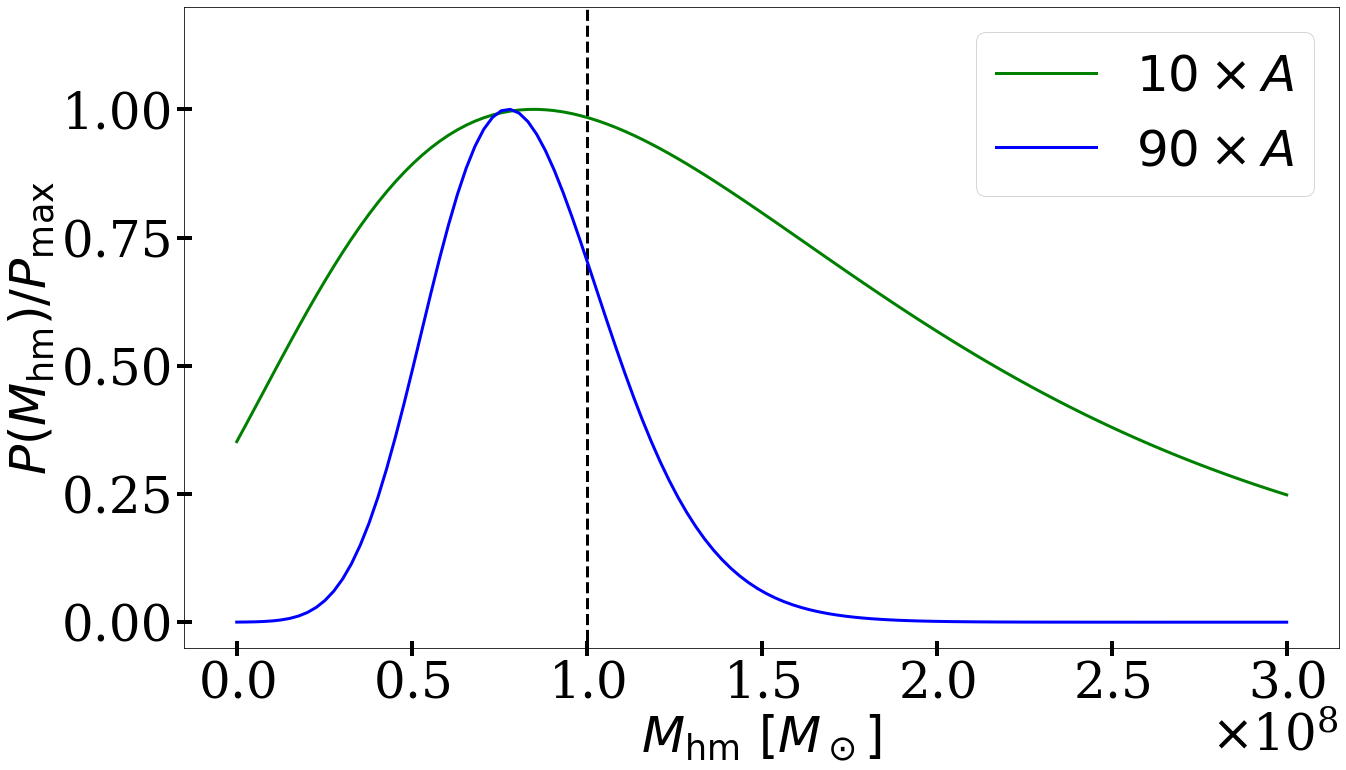}
    \caption{Likelihood distribution of half-mode mass for a \textit{mock} interloper population generated with $M_{\rm{hm}} = 10^8\mathrm{M}_\odot$ (dashed black line) for different numbers of future interloper detections that correspond to 10 and 90 times the effective line-of-sight area (A) of the JVAS B1938+666 system.}
    \label{fig:stat_fig2}
\end{figure}

An important prediction of WDM models is the suppression of structure below a characteristic mass scale due to the free streaming of dark matter particles in the early universe. This is quantified by the half mode mass tied to the particle mass of WDM with the scaling relation \citep{hm_mass_wdm_sch}
\begin{equation}\label{eq:scale}
    M_\mathrm{hm} \propto m^{-3.33}.
\end{equation}
We parametrize the effect of the half-mode mass on the halo mass function as
\begin{equation}
    n_\mathrm{WDM}(M) = n_\mathrm{CDM}(M) \left(1 + \frac{M_\mathrm{hm}}{M}\right)^{-1.3} ,
\end{equation}
which has been shown to match the results from N-body simulations \citep{hm_fit}. 
Just like in the last section, we divide the mass range $10^{8}-10^{11}\mathrm{M}_\odot$ into 30 logarithmic mass bins. We pick JVAS B1938+666 as a reference system and forecast the uncertainties on $M_\mathrm{hm}$ as the number of total detections increases. We populate the line-of-sight volume with interlopers following a mass function that has $M_\mathrm{hm} = 10^8 \mathrm{M}_\odot$, which is approximately the current upper bound from brightness fluctuations in quadruply lensed quasars \citep {wdm_chill,schechter}. In Fig. \ref{fig:stat_fig2}, we show the posterior probability distribution and 1$\sigma$ uncertainties for $8$ and $85$ interloper detections. These numbers are the expected number of interloper detections if we are able to probe an effective angular area in the sky that corresponds to 10 and 90 times the area of our reference system. With 8 detections, we are unable to constrain the half-mode mass significantly.
For 85 detections, we lower the uncertainty to $0.25 \times 10^8 \mathrm{M}_\odot$. This would allow us to rule out CDM at the 3$\sigma$ level.

\section{Results}
Using a Navarro-Frenk-White (NFW) profile \citep{NFW} with a free concentration parameter, we find that a model with an interloper at redshift $z_\mathrm{int}=1.42$ (the best fit), is preferred over a model with a subhalo ($z_\mathrm{int} = z_\mathrm{lens} = 0.881$) with a $\Delta(\mathrm{BIC}) = -17.2$ with one extra parameter, and a Bayes factor of $\log_{10}{\rm K}=3.4$, showing decisive evidence in favor of an interloper.
We run a dynamic nested sampling algorithm to obtain the posterior probability distributions for the model parameters. We find that the redshift of the perturber in JVAS B1938+666 is within $z_\mathrm{int} = 1.42\substack{+0.10 \\ -0.15}$, which confirms it is highly likely to be an interloper since the lens redshift of the same system is $z_\mathrm{lens} = 0.881$.

\subsection{Interloper Mass}
We find an interloper mass of  $M_{200} = 1.85\substack{+1.91 \\ -0.72}\times10^{9}\mathrm{M}_\odot$ for an NFW profile with a free concentration parameter. While this is an order of magnitude greater than the earlier result \citep{2012Natur} of $2.0 \times 10^8 \mathrm{M}_\odot$, this discrepancy can be explained by several differences in our analyses. Most importantly, the inferred mass of a perturber depends strongly on a truncation or cutoff radius \citep{despali2017}, which lies beyond the dense central region that lensing can effectively probe; this radius is thus never directly measured. The earlier analysis used the subhalo-specific effect of tidal stripping to estimate a truncation radius. However, we found the perturber highly likely to be an isolated interloper (any host large enough to tidally strip it would be luminous and directly visible), so the earlier analysis is not applicable. Since the NFW profile we use diverges, we quote an $M_{200}$ mass, following the convention of integrating mass out until the radius, $r_{200}$, within which the average halo density equals $200$ times the critical density of the universe. A second source of discrepancy is a known partial degeneracy between the interloper's mass and redshift \citep{mass_red_degen,despali2017}: since we found the interloper to lie at a higher redshift than the subhalo would be, it requires more mass to have the same lensing effect. 

We note that the best fit concentration found for the interloper (see Table \ref{table:2}) is well above the one expected following the mass-concentration relation \citep{mass_concen} (see Appendix \ref{sec:mass-conc} for further discussion). This has also been seen in previous work \citep{Minor:2020hic}, where a substructure system has been found with concentration at least 5$\sigma$ higher than the expected one from the mass-concentration relation. In fact, it has been pointed out \citep{Amorisco:2021iim} that detectable halos are very often high-concentration halos, and if one does not assume that all DM substructures lie on the mass-concentration relation, the $\Lambda$CDM expectation is, for many of the detectable DM perturbers in strong lenses, to be outliers.

Finally, we quantify the expected upper bound in the luminosity of the interloper by computing the magnification (and dimming) given its measured redshift, and find that an interloper that has the same brightness as a subhalo would appear to be roughly $0.52$ times as luminous, therefore changing the $3\sigma$ upper bound found for the same system \citep{2012Natur} to be $1\times10^8 L_\odot$.
By extrapolating the measured Faber-Jackson relation using the velocity dispersion we expect from this system, we estimate the luminosity to be roughly $2\times10^7 L_\odot$, well below the $3\sigma$ upper bound mentioned above. A higher-quality observation of the lensing system JVAS B1938+666 might enable us to measure the luminosity of the interloper. Such a follow up observation can be carried out by \textit{James Webb Space Telescope} (JWST) due to its high sensitivity and resolution in the near and mid infrared. It might be even possible to do a spectroscopic analysis on the emission of the interloper to measure its redshift independently of lensing.

\section{Conclusions}

As the sensitivity and scope of observations of strong gravitational lenses increase, it will become more and more feasible to measure whether a perturber lies in the main lens plane, especially if the perturber's redshift differs significantly from that of the main lens. As more perturbers are detected, it will become increasingly important to distinguish the line-of-sight halos from substructure. Confusing interlopers with subhalos gives inaccurate constraints on the subhalo mass function normalization and slope, which in turn affects the implications derived from those quantities on the nature of dark matter. 
In Fig.~\ref{fig:stat_fig1} we present a simulated analysis with the $x$-axis showing the slope of the subhalo mass function, and the $y$-axis displaying fraction of mass in substructures.
The blue cross denotes the ``true'' value used in a mock analysis, and the contours display the inferred values of the parameters if the interlopers are correctly identified or not.
Mischaracterizing interlopers as subhalos artificially boosts the total number of substructures found and, therefore, the inferred substructure mass fraction. 
Since many untested dark matter models predict suppression of structure in the sub-galactic mass scales that gravitational lensing can probe, this artificially high substructure mass fraction could lead to an incorrect falsification of such models. 

Moreover, the mass estimated for interlopers is significantly larger than the inferred total mass from tidal truncation (since the relevant cutoff radii are much larger than the tidal truncation radius). This means that a perturber that has been falsely assumed to be a subhalo is much more massive than it has been inferred to be. We might detect interlopers with masses of $\approx 10^8 \mathrm{M}_\odot$ that are mistakenly characterized as subhalos with masses of $\approx 10^7 \mathrm{M}_\odot$. If there is a suppression in substructure formation in that lower mass range, this false assumption would prevent us from probing it accurately. It would be interesting to more directly connect the central density profile measured by lensing to the properties of halos inferred from simulations under different dark matter scenarios without relying on the definition of the extended total mass used. We leave this for future work. 

There are complications such as baryonic physics and tidal stripping affecting the structure formation of subhalos. If we are able to accurately distinguish subhalos from interlopers, we can circumvent the complexities in tying the subhalo mass function to the nature of dark matter by instead focusing on interlopers and tying the halo mass function to dark matter models. As we probe lower mass scales with higher detection accuracy, we will be able to put tighter constraints on different dark matter models by measuring the number of interlopers, alleviating the complexities of baryonic feedback effects.
Moreover, deviations in the mass function of the interlopers from that of the subhalos might give us insights on the gravitational dynamics that govern the evolution of substructure under the influence of the potential of the main lens.

In summary, in this work we present the first dark perturber shown to be a line-of-sight halo through a gravitational lensing method. With tens of thousands of new lenses expected to become available in the near future, it will become even more important to distinguish between subhalos and line-of-sight halos to avoid obtaining wrong inferences about the nature of dark matter.

\section*{Acknowledgements}

We would like to thank Simon Birrer, Daniel Eisenstein, Anowar Shajib, and Sebastian Wagner-Carena for useful discussions and comments. CD is partially supported by the Department of Energy (DOE) Grant No. DE-SC0020223.
BO and AT are supported by the National Science Foundation under Cooperative Agreement PHY-2019786 (The NSF AI Institute for Artificial Intelligence and Fundamental Interactions, http://iaifi.org/).

\section*{Data Availability}

The system analyzed in this work is JVAS B1938+666. The data are publicly available on \url{https://mast.stsci.edu/portal/Mashup/Clients/Mast/Portal.html}. We analyze the image in this work using \texttt{lenstronomy} \citep{lenstronomy1,lenstronomy2}, a publicly available \texttt{Python} package for gravitational lensing. The code used for our analysis is available at \url{https://github.com/acagansengul/interlopers_with_lenstronomy}. 

\newpage
\bibliographystyle{mnras}
\bibliography{main} 

\newpage
\appendix

\section{Gravitational Lensing Formalism}

It is a well known prediction of General Relativity that a massive object bends the trajectory of light rays that pass nearby, in a phenomenon called \textit{gravitational lensing}. A gravitational lens is a collection of matter between an observer and a distant light source that distorts the light coming from the source. Galaxy-galaxy lensing is when both the background source that is distorted and the foreground lens that is causing the distortion are galaxies. The physical size of the lens is orders of magnitude smaller than the distances between the observer, the lens, and the source. Therefore, lensing at this scale can be well approximated by the \textit{thin-lens approximation}, where it is assumed that all the lens mass is concentrated on a single plane perpendicular to the line-of-sight. We can write a lens equation that maps the apparent angular position $\mathbf x \in \mathbb{R}^2$ of a point on the lens plane to its true angular position $\mathbf y \in \mathbb{R}^2$,
\begin{equation}
    \mathbf y = \mathbf x - \mathbf \alpha(\mathbf x),
\end{equation}
where $\mathbf \alpha \in \mathbb{R}^2$ is the \textit{deflection angle}. With the thin-lens approximation, the deflection angle is given by
\begin{equation}\label{alpha_eq}
    \mathbf{\alpha}(\mathbf x) = \dfrac{1}{\pi} \int_{\mathbb{R}^2}^{ } d^2 \mathbf x' \dfrac{\mathbf{x}-\mathbf{x'}}{|\mathbf{x}-\mathbf{x'}|^2} \kappa(\mathbf{x'}),    
\end{equation}
where $\kappa$ is the \textit{convergence} of the lens, defined as
\begin{equation}
    \kappa(\mathbf x) \equiv \frac{\Sigma(D_l\mathbf x)}{\Sigma_\mathrm{c}} \quad {\rm and} \quad \Sigma_\mathrm{c} \equiv \frac{c^2 D_s}{4\pi G D_l D_{ls}},
\end{equation}
where $\Sigma$ is the \textit{projected mass density}, $\Sigma_\mathrm{c}$ is the \textit{critical surface density} , $c$ is the speed of light, $G$ is the gravitational constant, $D_l$ and $D_s$ are the angular diameter distances to the lens plane and the source plane from the observer, respectively, and $D_{ls}$ is the angular diameter distance to the source plane from the lens plane. For a single lensing plane, the deflection angle can also be written as the gradient of a \textit{lensing potential} $\phi$ since it is a curl-free vector field:
\begin{equation}
    \mathbf \alpha = \nabla \phi.
\end{equation}
Convergence $\kappa$ and shear $\gamma$ can be written in terms of the second derivatives of the lensing potential:
\begin{equation}\label{eq:conv_and_shear}
    \kappa = \frac{1}{2}\left(\partial^2_1 \phi + \partial^2_2 \phi\right),\qquad \gamma_1 = \frac{1}{2}\left(\partial^2_1 \phi - \partial^2_2 \phi\right), \qquad \gamma_2 = \partial_1\partial_2\phi,
\end{equation}
where $\partial_1 \equiv \dfrac{\partial}{\partial x_1}$ and $\partial_2 \equiv \dfrac{\partial}{\partial x_2}$, with $\mathbf x = (x_1,x_2)$.

\section{Line-of-Sight Effects}\label{sec:line-of-sight}

Line-of-sight halos can have an effect on different lensing observables \citep{2010_Priya,McCully:2016yfe,despali2017,Gilman:2019vca}.
In particular, the light from a source galaxy is lensed by multiple deflectors along the line-of-sight. Since each perturber is still small in size compared to the cosmological distances, the thin-lens approximation still applies. The light gets lensed by a series of thin lensing sheets before it reaches the observer. Suppose we have two lens planes, denoted with $a$ and $b$, where $a$ is closer to the observer. The lens equations can be written as
\begin{eqnarray}\label{eq:lenseq2}
    \mathbf y &=& \mathbf x_a - \mathbf \alpha(\mathbf x_a), \qquad \mathbf \alpha(\mathbf x_a) = \mathbf \alpha_a(\mathbf x_a) + \mathbf \alpha_b(\mathbf x_b),\nonumber\\
    \mathbf x_b &=& \mathbf x_a - \Lambda_{ab}\mathbf \alpha_a(\mathbf x_a),\qquad \Lambda_{ab} \equiv \dfrac{D_{ab}D_s}{D_bD_{as}},
\end{eqnarray}
where $\mathbf x_a$ and $\mathbf x_b$ are the angular positions of the points where the light ray intersects with lens plane $a$ and $b$, respectively, $D_{ij}$ is the angular diameter distance between plane $i$ and plane $j$, and $D_i$ is the angular diameter distance between the observer and plane $i$. The index $s$ denotes the source plane. In this case, it is not possible to write a single lensing potential for the total angular deflection since the total angular deflection is not curl-free anymore. However, the angular deflection at each lens plane can still be written as the gradient of the lensing potential at that lens plane:
\begin{equation}\label{eq:gradients}
    \mathbf \alpha_a(\mathbf x_a) = \nabla_{\mathbf x_a}\phi_a, \qquad \mathbf \alpha_b(\mathbf x_b) = \nabla_{\mathbf x_b}\phi_b,
\end{equation}
where $\nabla_\mathbf{x_a}$ and $\nabla_\mathbf{x_b}$ are the gradient with respect to the coordinates on plane $a$ and $b$, respectively. The total angular deflection $\mathbf \alpha$ can be separated into $\mathbf \alpha_\mathrm{div}$ and $\mathbf \alpha_\mathrm{curl}$, a curl-free and a divergence-free component:
\begin{equation}
    \mathbf \alpha = \mathbf \alpha_\mathrm{div} + \mathbf \alpha_\mathrm{curl}, \qquad \nabla_{\mathbf x_a} \times \mathbf \alpha_\mathrm{div} = 0, \qquad \nabla_{\mathbf x_a} \cdot \mathbf \alpha_\mathrm{curl} = 0.
\end{equation}
To avoid confusion, we will call $\alpha_\mathrm{div}$ the divergence component and $\alpha_\mathrm{curl}$ the curl component. The divergence component can be fully produced using a single lensing plane with a mass distribution that produces a convergence,
\begin{equation}\label{eq:diveq}
    \kappa_\mathrm{div} = \frac{1}{2}\nabla_{\mathbf x_a} \cdot \mathbf \alpha=  \frac{1}{2}\nabla_{\mathbf x_a} \cdot \mathbf \alpha_\mathrm{div}.
\end{equation}
This equation can be inverted to calculate $\alpha_\mathrm{div}$, as shown in Eq. \eqref{alpha_eq}. 

The curl component is the signal that distinguishes line-of-sight effects from single-plane lensing \citep{int_pow_spec}. We can define an effective convergence for the curl component as
\begin{equation}\label{eq:curleq}
    \kappa_\mathrm{curl} \equiv \frac{1}{2} \nabla_\mathbf{x_a} \times \alpha = \frac{1}{2}\nabla_{\mathbf x_a} \times \mathbf \alpha_\mathrm{curl}.
\end{equation}
This equation can be inverted similar to the divergence case by
\begin{equation}\label{curl_inversion}
     \mathbf{\alpha_\mathrm{curl}}(\mathbf x) = \mathbf{\hat z} \times\dfrac{1}{\pi} \int_{\mathbb{R}^2}^{ } d^2 \mathbf x' \dfrac{\mathbf{x}-\mathbf{x'}}{|\mathbf{x}-\mathbf{x'}|^2} \kappa_\mathrm{curl}(\mathbf{x'}),    
\end{equation}
where $\mathbf{\hat z}$ is the unit vector that is orthogonal to the lens plane and points towards the observer.

We can express the curl component exactly in terms of the shear components of the mass distributions of lens plane $a$ and $b$. Substituting the gradients in Eq. \eqref{eq:gradients}
\begin{equation}
    \frac{1}{2}\nabla_{\mathbf x_a} \times \mathbf \alpha = \frac{1}{2}\nabla_{\mathbf x_a} \times \left[\alpha_a (\mathbf{x}_a) + \alpha_b (\mathbf{x}_b)\right] = \frac{1}{2} \nabla_{\mathbf x_a} \times \nabla_{\mathbf x_a} \phi_a + \frac{1}{2} \nabla_{\mathbf x_a} \times \nabla_{\mathbf x_b} \phi_b, 
\end{equation}
where the first term vanishes identically. Using Eq. \eqref{eq:lenseq2}, we calculate the partial derivatives and get
\begin{eqnarray}
    \frac{1}{2}\nabla_{\mathbf x_a} \times \mathbf \alpha &=& \Lambda_{ab} [(\partial_{b1}\partial_{b2}\phi_b)(\partial^2_{a2}\phi_a - \partial^2_{a1}\phi_a) \nonumber\\
    &+& (\partial_{a1}\partial_{a2}\phi_a)(\partial^2_{b1}\phi_b - \partial^2_{b2}\phi_b)],
\end{eqnarray}
where $\partial_{i1} \equiv \dfrac{\partial}{\partial x_{i1}}$ and $\partial_{i2} \equiv \dfrac{\partial}{\partial x_{i2}}$, with $\mathbf x_i = (x_{i1},x_{i2})$, with $i=a,b$. We can write the curl in a more compact form using Eq. \eqref{eq:conv_and_shear}:
\begin{equation}\label{eq:curl_comp}
    \kappa_\mathrm{curl} = \Lambda_{ab}\left[\gamma_{2a}(\mathbf x_a)\gamma_{1b}(\mathbf x_b) - \gamma_{1a}(\mathbf x_a)\gamma_{2b}(\mathbf x_b)\right].
\end{equation}

On top of capturing the line-of-sight effects of the interloper on lensing using the curl component, we also calculate the effective convergence of a 2-plane lensing system. We find that, for foreground interlopers, the corresponding effective subhalo can be placed as a direct projection from the observer onto the lens plane. For background interlopers, the corresponding effective subhalo needs to be located at the {\it apparent} position of the interloper, i.e. if the interloper were luminous, the position that it would appear to be at due to the lensing of the main lens,

\begin{eqnarray}
    \kappa_\mathrm{div} &=& \frac{1}{2}\nabla_{\mathbf x_a} \cdot \nabla_{\mathbf x_a}\phi_a + \frac{1}{2}\nabla_{\mathbf x_a} \cdot \nabla_{\mathbf x_b}\phi_b \\
    &=& \kappa_a(\mathbf x_a) + \kappa_b(\mathbf x_b) + \Lambda_{ab}[\gamma_{1a}(\mathbf x_a)\gamma_{1b}(\mathbf x_b) \nonumber\\
    &-& \gamma_{2a}(\mathbf x_a)\gamma_{2b}(\mathbf x_b) - \kappa_{a}(\mathbf x_a)\kappa_{b}(\mathbf x_b)]. \label{eq:divcoup}
\end{eqnarray}

We find that it is much more practical to use {\it apparent} positions $\mathbf{x}_b$ as the model parameters \citep{DES:2019fny} for the interloper redshift for a background interloper. This is because the true angular position for a background interloper gets lensed by the main lens, which results in a strong correlation with interloper redshift. 

\begin{figure}
    \centering
    \includegraphics[width=\linewidth]{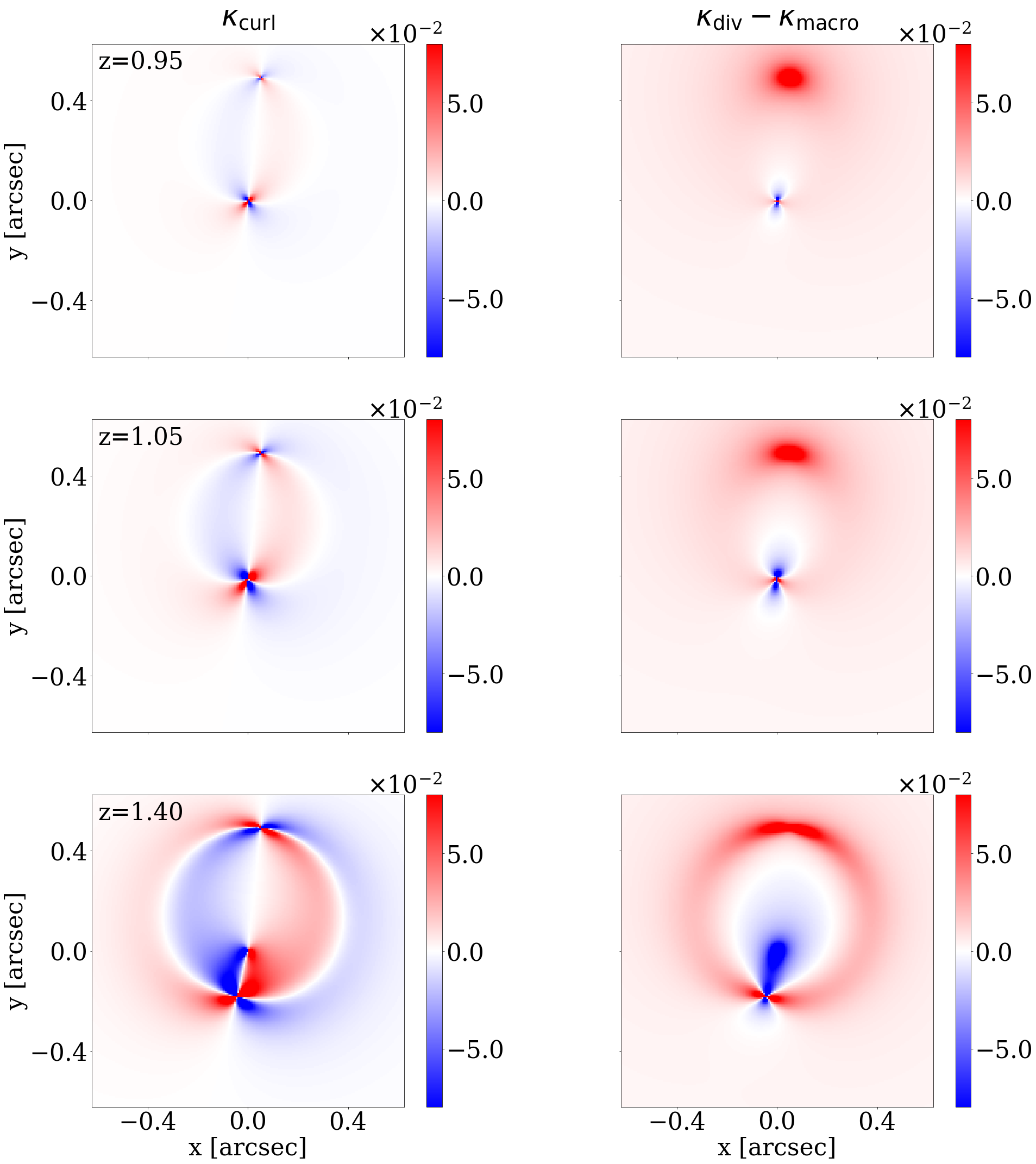}
    \caption{\textit{Left column:}  $\kappa_\mathrm{curl}$ given by Eq. \eqref{eq:curleq}. \textit{Right column:} $\kappa_\mathrm{div}$ given by Eq. \eqref{eq:diveq} with the main lens convergence $\kappa_\mathrm{macro}$ subtracted, for a lensing system with a main lens with an elliptical power law mass profile at $z=0.881$, a circularized Einstein radius of $\theta_E = 0.46''$, a negative power law slope of $\gamma = 2.30$ , eccentricity components $e_1 = -0.03, e_2 = 0.1$, and with a singular isothermal sphere interloper at $z_\mathrm{int}=0.95, 1.05,$ and $1.40$ for \textit{top},\textit{ middle}, and \textit{bottom} rows, respectively, with an Einstein radius of $b_\mathrm{int}=12.5$ milliarcseconds and an apparent position of $[0.05,0.48]$ arcseconds. These parameters are chosen to be similar to the real lensing system JVAS B1938+666 and its detected perturber. We see that, as the redshift difference between the interloper and the main lens increases, the curl component gets stronger and the effective convergence becomes elongated.}   
    \label{fig:curl}
\end{figure}

The curl component shown in Eq. \eqref{eq:curl_comp}, together with the third, fourth and fifth terms in Eq. \eqref{eq:divcoup}, causes deviations in the angular deflections of the multi-plane lensing system from a purely single plane one. The shift in pixel brightness caused by this change in angular deflections is what is fit to constrain the redshift of the perturber. This line-of-sight effect becomes stronger as the perturber gets further away from the main lens plane, which can be seen in Fig. \ref{fig:curl}.

\section{Mass-Concentration Relation}
\label{sec:mass-conc}

\begin{figure}
    \centering
    \includegraphics[width=\linewidth]{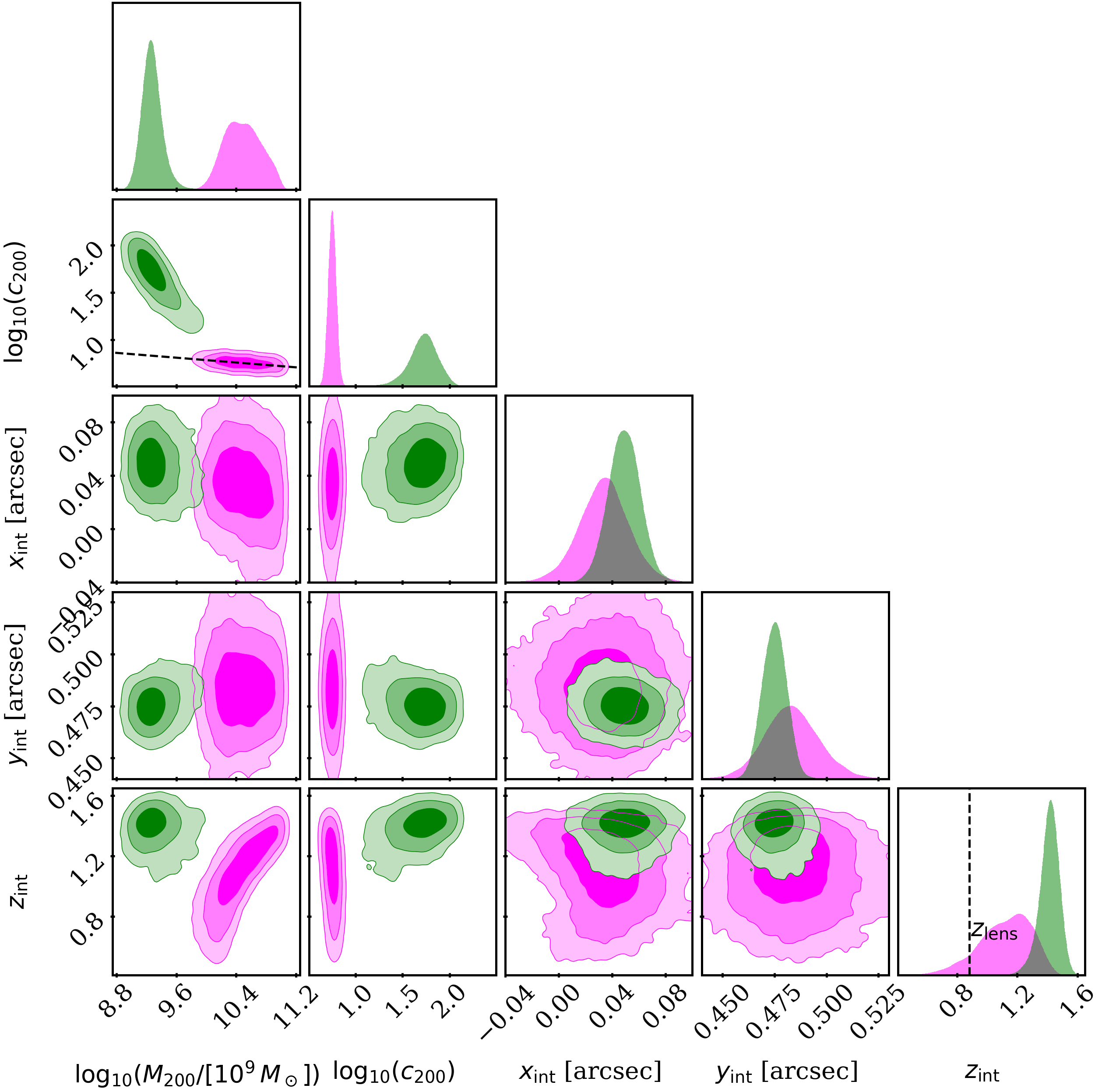}
    \caption{Posterior probability distributions of the interloper parameters in the real data modeled as an NFW profile with a free concentration parameter (green), and an NFW profile that follows a mass-concentration relation (magenta) obtained from CDM simulations \citep{mass_concen}, given by Eq. \eqref{eq:m200c200}. The mass-concentration relation is shown as a horizontal black dashed line. The scatter in $c_{200}$ around the dashed line is due to the scatter in redshift and mass of the models considered.}  
    \label{fig:mass_concen}
\end{figure}

In addition to the models described in Table \ref{table:1}, we ran a model with an NFW interloper where we impose the following mass-concentration relation \citep{mass_concen},
\begin{equation}\label{eq:m200c200}
    \log_{10}(c_{200},z) = b(z)\log_{10}(M_{200}/10^{12} \mathrm{M}_\odot h^{-1}) + a(z),
\end{equation}
where
\begin{align}
    b(z) &= -0.101 + 0.026 \,z,\\
    a(z) &= 0.520 + (0.905 - 0.520)\exp[-0.617\,z^{1.21}],
\end{align}
which is obtained from simulations. We will call this model $M_\mathrm{intNFWc10}$. In Fig. \ref{fig:mass_concen} we show the posteriors of this model compared to $M_\mathrm{intNFW10}$, where the concentration is allowed to vary freely. The relation given by Eq. \eqref{eq:m200c200} is shown as the black dashed line. When $M_{200}$ and $c_{200}$ are freely varied, they show a strong inverse correlation (see Fig. \ref{fig:nfws}). Since Eq. \eqref{eq:m200c200} predicts $c_{200} \approx 10$ in the mass and redshift range of the interloper, the inverse correlation mentioned earlier results in a much higher mass prediction of $M_{200}\approx 3\times 10^{10}$. However, $M_\mathrm{intNFW10}$ is preferred over $M_\mathrm{intNFWc10}$ with a Bayes factor of $\log_{10}{\rm K} = 2.9$, showing decisive evidence in favor of a model with a concentration parameter that is higher than the one expected from the mass-concentration relation.

\label{lastpage}
\end{document}